\documentclass[english,12pt,aps,prd,a4paper,preprintnumbers,floatfix,nofootinbib 
,superscriptaddress, notitlepage]{revtex4} 
 \pdfoutput=1
\usepackage[usenames,dvipsnames]{color}  
\usepackage{graphicx}
\usepackage{setspace}
\usepackage{caption}
\usepackage{amsmath}
\usepackage{amssymb}
\usepackage[colorlinks=true,citecolor=darkred,urlcolor=darkred, pdfborder={0 0 0}]{hyperref}
\usepackage[normalem]{ulem}
\usepackage{xcolor}
\usepackage{color}
\makeatletter
\def\p@subsection{}
\makeatother

\definecolor{darkred}{rgb}{0.6,0,0}

\definecolor{linkcolor}{rgb}{0,0,0.5}

\def\gsim{\raise0.3ex\hbox{$\;>$\kern-0.75em\raise-1.1ex\hbox{$\sim\;$}}}
\def\lsim{\raise0.3ex\hbox{$\;<$\kern-0.75em\raise-1.1ex\hbox{$\sim\;$}}}

\def\beqn#1{\begin{equation}\label{#1}}
\def\eeqn{\end{equation}}

\def\beqa#1{\begin{eqnarray}\label{#1}}
\def\eeqa{\end{eqnarray}}

\def\znbb {neutrinoless double beta decay }

\def\Z2{$\mathcal{Z_2}$}

\newcommand {\ignore}[1]{}

\def\lfv{lepton flavour violation }
\def\lnv{lepton number violation }

\def\clfv{lepton flavour violation }
\def\lnvg{lepton number violating }
\def\SM{$\mathrm{SU(3)_c \otimes SU(2)_L \otimes U(1)_Y}$ }

\def\321{$\mathrm{SU(3) \otimes SU(2) \otimes U(1)}$ }

\newcommand{\no}{\nonumber\\}

\newcommand{\AddrAHEP}{
  AHEP Group, Institut de F\'{i}sica Corpuscular --
  CSIC/Universitat de Val\`{e}ncia, Parc Cient\'ific de Paterna.\\
 C/ Catedr\'atico Jos\'e Beltr\'an, 2 E-46980 Paterna (Valencia) - SPAIN}
 
 \newcommand{\AddrIISERB}{Department of Physics, Indian Institute of Science Education and Research - Bhopal, \\ 
Bhopal Bypass Road, Bhauri, Bhopal 462066, India}

\begin{document}

\title{\color{BrickRed} W-mass Anomaly in the Simplest Linear Seesaw Mechanism}
\author{Aditya Batra}\email{adityab17@iiserb.ac.in}
\affiliation{\AddrIISERB}
\author{ Praveen Bharadwaj}\email{praveen20@iiserb.ac.in}
\affiliation{\AddrIISERB}
\author{Sanjoy Mandal}\email{smandal@kias.re.kr}
\affiliation{Korea Institute for Advanced Study, Seoul 02455, Korea}
\author{Rahul Srivastava}\email{rahul@iiserb.ac.in}
\affiliation{\AddrIISERB}
\author{Jos\'{e} W. F. Valle}\email{valle@ific.uv.es}
\affiliation{\AddrAHEP}

\begin{abstract}
  \vspace{1cm} 
  
  The simplest linear seesaw mechanism can accommodate the new CDF-II $W$ mass measurement. In addition to Standard Model particles, the model includes quasi-Dirac leptons, and a second, leptophilic, scalar doublet seeding small neutrino masses. Our proposal is consistent with electroweak precision tests, neutrino physics, rare decays and collider restrictions, requiring a new charged scalar below a few TeV, split in mass from the new degenerate scalar and pseudoscalar neutral Higgs bosons.
   
\end{abstract}
\maketitle

\section{Introduction}
\label{sec:introduction}

Progress in high energy physics often means discovering new fundamental particles.
Ten years ago we had a vivid example of this with the historic discovery of the Higgs boson by the ATLAS~\cite{Aad:2012tfa} and CMS~\cite{Chatrchyan:2012ufa}
experiments at CERN, which reassured that the particle physicists' view on electroweak symmetry breaking is correct.
However, precision measurements of key particles, such as neutrinos and the W boson, could be equally important.
Non-zero neutrino masses~\cite{Kajita:2016cak,McDonald:2016ixn,KamLAND:2002uet,K2K:2002icj} constitute one of the most convincing proofs of new physics.
Unfortunately, despite many attempts, the challenge of underpinning the origin of neutrino mass remains as open as ever. 
Recently, the CDF-II collaboration published their high precision measurement of the $W$ boson mass $m_{W}^{\rm CDF} = 80.4335 \pm 0.0094$ GeV~\cite{CDF:2022hxs},
which reveals a $7$-$\sigma$ difference from the Standard Model (SM) expectation $m_{W}^{\rm SM} = 80.354 \pm 0.007$ GeV~\cite{Zyla:2020zbs},
which seemingly cannot be explained solely within the SM framework, e.g. by uncertainties in parton distribution functions~\cite{Isaacson:2022rts,Gao:2022wxk} .
If confirmed, this measurement could reveal cracks in our basic understanding of particle physics.
To account for this large deviation from the SM $W$-mass prediction, novel physics beyond the Standard Model (BSM) seems required. 

Even though the SM lacks a neutrino mass one can introduce it effectively through a dimension-five lepton number non-conserving operator~\cite{Weinberg:1980bf}. 
\begin{align}
-\mathcal{L}_\nu^{d=5}=\frac{1}{\Lambda}(\bar{L}_L \Phi)\, (\Phi^{T}L_L^c) + \text{H.c.},
\end{align} 
where $\Lambda$ is some effective (presumably large) mass scale, $\Phi$ is the SM Higgs doublet and $L_{L}$ is the lepton doublet. 
This leads to Majorana neutrino masses below the electroweak breaking scale.  
 The seesaw mechanism provides a specially interesting UV-completion of this operator, most generally realized in the simplest \SM gauge structure~\cite{Schechter:1980gr}.
 In its type-I realization neutrinos get mass due to the exchange of heavy singlet mediators. 
 The left-handed neutrinos acquire a mass as $m_\nu\sim m_D^2/M_N$, with $m_D=Y_\nu v/\sqrt{2}$, so that $\Lambda=M_N/Y_{\nu}^2$. 
 Hence, for $m_\nu\sim\mathcal{O}(0.1\,\text{eV})$ and relatively large Yukawa coupling $Y_\nu\sim\mathcal{O}(1)$, $M_N$ must be large, i.e., $M_N\gg\mathcal{O}(\text{TeV})$.
 As a result, such high-scale implementation of the seesaw mechanism has few phenomenological implications beyond those related to neutrino masses. 

 However, the seesaw paradigm can also arise from low-scale physics~\cite{Boucenna:2014zba}. 
 Two varieties of low-scale seesaw mechanism share a common ``(3,6)'' template~\cite{Schechter:1980gr,Schechter:1981cv}~\footnote{These contain 6 singlets that make up 3 heavy Dirac leptons in the limit of lepton number conservation. } which,
 instead of adding a single right-handed neutrino, requires a pair of isosinglets associated to each family. 
 These varieties are the inverse~\cite{Mohapatra:1986bd,Gonzalez-Garcia:1988okv} and the linear seesaw mechanisms~\cite{Akhmedov:1995ip,Akhmedov:1995vm,Malinsky:2005bi}. 
 In both schemes, taking the lepton-number-conserving limit, leads to a 
 massless neutrino setup that forms the template for building a genuine low-scale seesaw mechanism, in which the small neutrino masses are mediated by
 heavy quasi-Dirac neutrinos. These might be accessible at high energies~\cite{Dittmar:1989yg,Gonzalez-Garcia:1990sbd,AguilarSaavedra:2012fu,Das:2012ii,Deppisch:2013cya},
 a possibility taken up by the ATLAS and CMS collaborations at the LHC~\cite{ATLAS:2019kpx,CMS:2018iaf,CMS:2022nty},
 and which is also in the agenda of future proposals, see e.g.~\cite{Drewes:2019fou,Abdullahi:2022jlv}.

\par  Prompted by the CDF-II high precision measurement of the $W$ boson mass we examine the simplest SM-based variant of the linear seesaw mechanism.
In contrast to most previous formulations~\cite{Akhmedov:1995ip,Akhmedov:1995vm,Malinsky:2005bi}, here we do not impose left-right symmetry.  
Namely, the linear seesaw mechanism is realized within the \SM gauge structure itself, in which lepton number symmetry is ungauged~\cite{Fontes:2019uld}.  
We assume at least two pairs of isosinglet leptons~\footnote{If we have only two pairs, instead of three, the \znbb rate has a lower bound even for normal neutrino mass
  ordering~\cite{Reig:2018ztc,Barreiros:2018bju}, and should be soon detectable for inverse mass ordering~\cite{Mandal:2019oth,Avila:2019hhv}.}.
In addition, a key feature of the model is that it requires a second Higgs doublet, carrying two units of lepton number. 

Even for TeV-scale mediators, neutrino masses are small due to the small VEV of this second Higgs doublet. 
In contrast to~\cite{Fontes:2019uld} we assume that lepton number symmetry is broken explicitly, but softly, in the scalar potential.
This way we avoid a Nambu-Goldstone boson and the associated stringent astrophysical restrictions. 
This results in a motivated version of the two-doublet Higgs model~\cite{Grimus:2007if,Grimus:2008nb,Branco:2011iw,Bhattacharyya:2015nca,Wang:2022yhm},
allowing for possible experimental tests of its validity.

\par Here we show that the CDF-II high precision $W$ boson mass measurement~\cite{CDF:2022hxs} at odds with the SM expectation can be easily accommodated. 
In order to account for this large deviation from the SM $W$-mass prediction, new physics seems required. An interesting example is provided by neutrino mass generation schemes, which often
 require new particles that may shift the W-boson mass appreciably at the radiative level~\cite{Borah:2022obi,Popov:2022ldh,Batra:2022org,Batra:2022pej,CentellesChulia:2022vpz,Chakrabarty:2022voz,VanLoi:2022eir}.

In this work we show that, despite its simplicity, the linear seesaw can easily account for the CDF-II anomaly in the $W$ boson mass measurement,
through the new contributions from the extra scalars in the loop corrections of the gauge boson self energies.  
These two-point functions are parameterised through the so-called oblique parameters $S$, $T$ and $U$.   
We show that the new CDF-II $W$ boson mass measurement can be consistent with these oblique parameters, as well as with perturbative unitarity and vacuum stability in a consistent manner. 
We describe the implications for the Higgs boson mass spectrum arising from the CDF II result as well as neutrino physics and theoretical consistency.

\par The paper is organized as follows. In Sec.~\ref{sec:model} we briefly describe the model, giving the details of the new fields and their interactions.
In Sec.~\ref{sec:stu}, we examine the $S$, $T$, and $U$ parameter space allowed by the new CDF-II results, along with the other theoretical and phenomenological constraints. 
  Finally, in Sec.~\ref{sec:conclusions} we conclude.

\section{Linear seesaw model} 
\label{sec:model}

This low-scale seesaw variant was first proposed within a $SU(3)_c\otimes SU(2)_L\otimes SU(2)_R\otimes U(1)_{B-L}$ setup~\cite{Akhmedov:1995ip,Akhmedov:1995vm},
and subsequently shown to arise naturally in the $SO(10)$ framework in the presence of gauge singlets~\cite{Malinsky:2005bi}. 
Here we focus on the simplest effective variant of the linear seesaw mechanism, realized within the \SM gauge structure. 
In addition to the SM Higgs scalar $\Phi$, we add one more doublet $\chi_L$. 
In contrast with Ref.~\cite{Fontes:2019uld} we assume an explicit breaking of lepton number symmetry.
This avoids the existence of a physical (nearly) massless Nambu-Goldstone boson, and the associated restrictions from LEP~\cite{Joshipura:1992hp}
as well as the stringent stringent astrophysical limits from stellar cooling~\cite{Fontes:2019uld}.
The additional doublet scalar $\chi_L$ carries lepton number $L[\chi_L]=-2$.   
In addition we have the usual lepton singlets $\nu_i^c$ with lepton number $L[\nu^c_i]=-1$ and lepton singlets $S_i$ with lepton number $L[S_i]=1$. 
Lepton number is a global $U(1)$ symmetry broken in the scalar sector, explicitly but softly.  
\subsection{Neutrino mass generation}
\label{subsec:neutrino-mass}

Our simplest linear seesaw setup lies closer to the Standard Model. 
The relevant lepton-number-invariant Lagrangian for neutrino mass generation is written as 
\begin{equation}
  \label{eq:Yukawa}
  - \mathcal{L}_{\rm Yuk}= Y_{\nu}^{ij} L_i^T C \nu^c_j \Phi 
  + M_R^{ij} \nu^c_i C S_j + Y_{S}^{ij} L_i^T C  S_j \chi_L+ \text{h.c.}
\end{equation}
where $Y_{\nu}$ and $Y_{S}$ are dimensionless Yukawa couplings, $M_{R}$ is an arbitrary bare mass term, $\Phi$ is the SM Higgs doublet, and $\chi_L$ is the second scalar doublet. 
This form represents an effective description of more complete UV-completions of linear seesaw with gauged lepton number~\cite{Akhmedov:1995ip,Akhmedov:1995vm,Malinsky:2005bi}. 
\begin{figure}[!htbp]
	\includegraphics[height=4cm,width=0.7\textwidth]{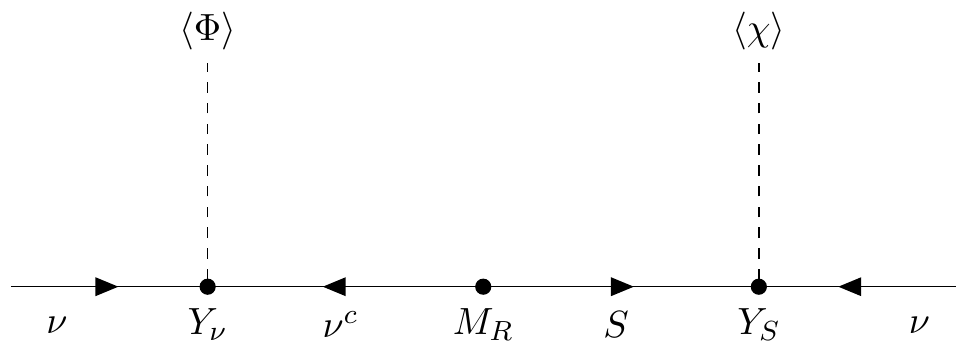}
	\caption{The symmetrized diagram illustrates neutrino mass generation in the linear seesaw. }
	\label{fig:neutrino}
\end{figure}

After \lnv it will give the linear seesaw mass matrix in the basis $\nu,\nu^c,S$ as follows  
\begin{align}
\mathcal{M}_{\nu}=
 \begin{pmatrix}
  0 & m_D  & M_L  \\
  m_D^T & 0 & M_R \\
  M_L^T &  M_R^T  &  0  \\
 \end{pmatrix}
 \label{eq:neutrino-mass-matrix}
\end{align}
where $M_R$ is a bare mass, $m_D=\frac{Y_\nu^{ij}v_{\Phi}}{\sqrt{2}}$, $M_L=\frac{Y_S^{ij}v_{\chi}}{\sqrt{2}}$ are determined by the VEVs of the $\Phi$ and $\chi_L$ doublets, $v_\Phi$ and $v_\chi$, respectively. 
%
The full neutrino mass matrix in Eq.~\eqref{eq:neutrino-mass-matrix} is diagonalized by a unitary matrix of the same dimension as: 
\begin{align}
\mathcal{U}_0^T \mathcal{M}_{\nu} \mathcal{U}_0=\mathcal{M}_\nu^{\rm diag}.
\end{align} 
Following the standard procedure of two-step diagonalization $\mathcal{U}_0$ can be expressed as~\cite{Schechter:1981cv}, 
\begin{align}
\mathcal{U}_0=\mathcal{U} V\,\,\text{ with }\,\, \mathcal{U}^T \mathcal{M}_\nu \mathcal{U}=\begin{pmatrix} m_{\rm light} & 0 \\
0 & M_{\rm heavy}\\
\end{pmatrix}
\end{align}
Hence, $\mathcal{U}$ is the matrix which brings the full neutrino matrix to block diagonalized form, while $V=\text{diag}(U_{\rm lep}, V_2)$ diagonalizes the mass matrices in the light and heavy sector.
One can approximately write the matrix $\mathcal{U}$ as follows 
\begin{eqnarray}\label{u-bdiag}
&\mathcal{U}\approx
\left(
\begin{array}{ccc}
I & 0 & 0\\
0 & \frac{1}{\sqrt{2}}I & -\frac{1}{\sqrt{2}}I\\
0 & \frac{1}{\sqrt{2}}I & \frac{1}{\sqrt{2}}I
\end{array}
\right)
\left(
\begin{array}{ccc}
I-\frac{1}{2}S_1\,S_1^\dagger & 0 & iS_1\\
0 & I & 0\\
iS_1^\dagger & 0 & I-\frac{1}{2}S_1^\dagger \,S_1
\end{array}
\right)
\left(
\begin{array}{ccc}
I-\frac{1}{2}S_2\,S_2^\dagger & iS_2 & 0\\
iS_2\dagger & I-\frac{1}{2}S_2^\dagger \,S_2 & 0\\
0 & 0 & I
\end{array}
\right),&\nonumber\\
&
\approx
\left(
\begin{array}{ccc}
I & iS_2 & iS_1\\
-i\frac{1}{\sqrt{2}}\left(S_1^\dagger-S_2^\dagger \right) & \frac{1}{\sqrt{2}}I & -\frac{1}{\sqrt{2}}I\\
 i\frac{1}{\sqrt{2}}\left(S_1^\dagger+S_2^\dagger \right)& \frac{1}{\sqrt{2}}I & \frac{1}{\sqrt{2}}I
\end{array}
\right)+O(\epsilon^2)&
\end{eqnarray}
where $S_1$ and $S_2$ are $3\times 3$ matrices. In the limit $M_L \to 0$ we have  $S_1=S_2=S$ where
\begin{equation} \label{eq:mixing}
iS^*=-m_D \,(M_R^T)^{-1} \sim \epsilon 
\end{equation} 
governs the seesaw diagonalization procedure~\cite{Schechter:1981cv}, and is small, due to the hierarchy $M_R\gg m_D\gg M_L$.
Eq.~\eqref{eq:neutrino-mass-matrix} leads to the effective light neutrino mass matrix, i.e.
\begin{equation}\label{lin}
m_{\rm light}=m_D(M_L M_R^{-1})^T+(M_L M_R^{-1}){m_D}^T.
\end{equation}

In the limit $M_L\to 0$ it leads to three massless neutrinos as in the SM.
In contrast to the SM, however, neutrinos mix with each other~\cite{Valle:1987gv} and with the heavy Dirac leptons, $N$.
The resulting forms for the charged and neutral current weak interactions are given, for example, in~\cite{Bernabeu:1987gr,Branco:1989bn,Rius:1989gk}.
This massless limit is the template upon which all low-scale seesaw schemes are built, including also the inverse seesaw mechanism~\cite{Mohapatra:1986bd,Gonzalez-Garcia:1988okv}. 
Turning on non-zero $M_L$ leads to three light Majorana eigenstates $\nu_i$, with $i=1,2,3$ in addition to the heavy ones $N_j$.  

The Feynman diagram in Fig. \ref{fig:neutrino} helps to understand the neutrino mass generation mechanism. 
In contrast to conventional type-I seesaw setups, the matrix $m_{\rm light}$ scales linearly with the Dirac Yukawa couplings contained in $m_D$, hence the name linear seesaw mechanism. 
Neutrino masses are suppressed by the small factor $M_L/M_R$ irrespective of how low is the $M_R$ scale characterizing the heavy messengers. 
This suppression requires a small value of $v_\chi$. In the limit $v_\chi\to 0$ lepton number is restored, so the construction is symmetry-protected.

\begin{figure}[!htbp]
	\centering
	\includegraphics[height=6.5cm,width=0.65\textwidth]{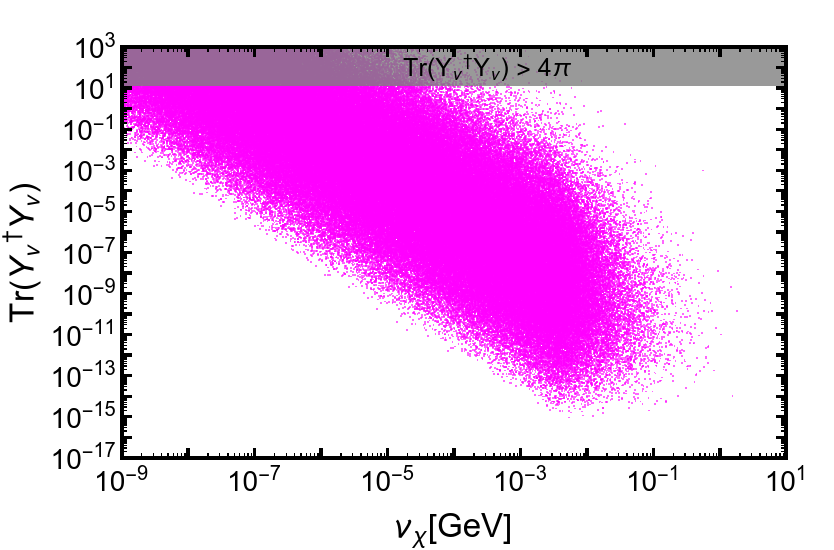}
	\caption{
          $\text{Tr}(Y_\nu^\dagger Y_\nu)$ consistent with neutrino masses versus the \lnvg VEV $v_\chi$, Eq.~\eqref{eq:Ynu}.
          We took $m_1\leq 0.1$~eV and set CP phases to zero, fixing oscillation parameters to their best fit values~\cite{deSalas:2020pgw,10.5281/zenodo.4726908}.
          The grey band violates perturbativity. 
   We varied the other parameters as
   $1~\text{TeV}\leq M_R^{ii}\leq 100~\text{TeV}$, $10^{-4}\leq Y_{S}^{ij}\leq \sqrt{4\pi}$ with $|M_L|\leq 10^{-1}$ GeV.   
   }
	\label{fig:vchi}
\end{figure}

The Dirac Yukawa couplings can be expressed in terms of the oscillation parameters in the following form~\cite{Forero:2011pc,Cordero-Carrion:2018xre,Cordero-Carrion:2019qtu}, 
\begin{align}
Y_\nu=\frac{\sqrt{2}}{v_\Phi}U_{\rm lep} \text{diag}\{\sqrt{m_i}\} \, A^T \text{diag}\{\sqrt{m_i}\} \, U_{\rm lep}^T \left(M_L^T\right)^{-1} \,M_R^T,
\label{eq:Ynu}
\end{align}
where $U_{\rm lep}$ is approximately the mixing matrix determined in oscillation experiments, $m_i$ are the three light neutrino masses and $A$ has the following general form: 
\begin{equation}\label{A}
\left(
\begin{array}{ccc}
\frac{1}{2} & a & b\\
-a & \frac{1}{2}& c\\
-b & -c  & \frac{1}{2}
\end{array}
\right),
\end{equation}
with $a,b,c$ real numbers. In this case $M_L$ and $M_R$ are general real matrices. One can always go to a basis where $M_R$ is diagonal.
As we will see later, this generalized Casas-Ibarra parametrization is particularly useful for our purposes.
Following Eq.~\eqref{eq:Ynu}, the values of $v_\chi$ consistent with neutrino mass are shown in Fig.~\ref{fig:vchi}.

\par 
One can see that large Yukawa couplings can lead to adequately small values of the VEV $v_\chi$ characterizing \lnv.
Here we took the lightest neutrino mass in the range $10^{-5}\,\text{eV}-0.1$~eV, with diagonal $M_R$ varying from 1 TeV to 100 TeV.
We have randomly generated various entries of $Y_S$ matrix in the range $10^{-4}-\sqrt{4\pi}$ such that $|M_L|\leq 10^{-1}$ GeV.
We also varied the off-diagonal elements of $A$ matrix in the range $0-10^{-2}$.
The oscillation parameters were fixed to their best fit values~\cite{deSalas:2020pgw,10.5281/zenodo.4726908} and CP phases were neglected.

\subsection{The scalar sector} 
\label{subsec:scalar-sector}

Apart from the Standard Model Higgs doublet $\Phi$ we have a second scalar doublet $\chi_L$ charged under lepton number. 
The \SM gauge invariant scalar potential is given by  
\begin{align}
 V&=-\mu_\Phi^2 \Phi^{\dagger}\Phi - \mu_\chi^2 \chi_L^{\dagger}\chi_L+\lambda_1 (\Phi^{\dagger}\Phi)^2 + \lambda_2 (\chi_L^{\dagger}\chi_L)^2+\lambda_3 \chi_L^{\dagger}\chi_L \Phi^{\dagger}\Phi\nonumber \\
 & + \lambda_4 \chi_L^{\dagger}\Phi \Phi^{\dagger}\chi_L - \left(\mu_{12}^2 \Phi^{\dagger}\chi_L+H.c.\right),
 \label{eq:potential}
\end{align}
where the soft $\mu_{12}^2$ term breaks explicitly lepton number symmetry inducing a small VEV $v_\chi \propto \mu_{12}$. 
For definiteness, in the above, we assumed all parameters to be real. 

We now examine the consistency conditions of the potential. 
To ensure that the scalar potential is bounded from below and has a stable vacuum at any given energy scale, the following constraints must hold:  
\begin{align}
\lambda_1\geq 0,\,\, \lambda_2 \geq 0,\,\, \lambda_3 \geq -2\sqrt{\lambda_1 \lambda_2}\,\, \text{and}\,\, \lambda_3 + \lambda_4 \geq -2\sqrt{\lambda_1 \lambda_2}.
\label{vacstab}
\end{align}
To ensure perturbativity, we restricted the scalar quartic couplings in Eq.~\ref{eq:potential} as $\lambda_i\leq 4\pi$.

\subsection*{Higgs boson mass spectrum}


In order to obtain the mass spectrum for the scalars after \SM and lepton-number symmetry breaking, we expand the doublet scalar fields $\Phi$ and $\chi_L$ as  
\begin{align}
 \Phi=
 \begin{pmatrix}
  \Phi^{+} \\
   \frac{1}{\sqrt{2}}(v_\Phi+h_\Phi+i\eta_\Phi)\\
 \end{pmatrix},\hspace{1cm}  \chi_L=
 \begin{pmatrix}
  \chi^{+} \\
   \frac{1}{\sqrt{2}}(v_\chi+h_\chi+i\eta_\chi)\\
 \end{pmatrix}
\end{align}
where $h_\Phi$, $h_\chi$ and $\eta_\Phi$, $\eta_\chi$ are CP even and CP odd neutral scalars, while $\chi^{\pm}$ and $\Phi^{\pm}$ are four charged scalars.
In order to get the physical states we need to diagonalize the charged and neutral scalar mass matrices
taking into account the allowed mixing between the two doublets $\Phi$ and $\chi_L$ in the above scalar potential. 

Besides the three unphysical Goldstone bosons $G^\pm,G^0$, which are ``eaten'' to become the longitudinal components of the SM $W^{\pm}$ and $Z$ gauge bosons,
there are five physical mass eigenstates $h$, $H$, $H^{\pm}$, $A$.  
The mass matrix for the charged scalars in the basis $(\chi^+,\,\,\Phi^{+})$ is 
\begin{align}
 \mathcal{M}^2=
 \begin{pmatrix}
 \mu_{12}^2 \frac{v_\Phi}{v_\chi}-\lambda_4\frac{v_\Phi^2}{2}     &  -\mu_{12}^2+\lambda_4\frac{v_\Phi v_\chi}{2} \\
  -\mu_{12}^2+\lambda_4\frac{v_\Phi v_\chi}{2} &   \mu_{12}^2\frac{v_\chi}{v_\Phi}-\lambda_4\frac{v_\chi^2}{2} \\
 \end{pmatrix}
\end{align}
One can easily see that this matrix has a zero eigenvalue corresponding to the charged Goldstone boson $G^{+}$, and that the physical charged Higgs has a mass 
\begin{align}
 m_{H^{\pm}}^2=v^2\left(\frac{\mu_{12}^2}{v_\Phi v_\chi}-\frac{\lambda_4}{2}\right)
\end{align}
where $v=\sqrt{v_\Phi^2+v_\chi^2}$. The mass eigenstates are obtained as 
\begin{align}
\begin{pmatrix}
\chi^+ \\
\Phi^+
\end{pmatrix}=R(\beta)\begin{pmatrix}
G^+\\
H^+
\end{pmatrix}=\begin{pmatrix}
\cos\beta & -\sin\beta \\
\sin\beta &  \cos\beta 
\end{pmatrix}\begin{pmatrix}
G^+\\
H^+
\end{pmatrix}\,\,
\text{with} \,\, \tan\beta=\frac{v_\Phi}{v_\chi}
\end{align}

The mass matrix for CP even neutral scalars in the basis $(h_\chi\,\,h_\Phi)$ is given as 
\begin{align}
 \mathcal{M}_{h}^2=
\begin{pmatrix}
 A & C \\
 C  & B \\
\end{pmatrix}=
\begin{pmatrix}
 \mu_{12}^2\frac{v_\Phi}{v_\chi}+2\lambda_2 v_\chi^2  & -\mu_{12}^2+v_\Phi v_\chi \lambda_{34} \\
-\mu_{12}^2+v_\Phi v_\chi \lambda_{34} &  \mu_{12}^2\frac{v_\chi}{v_\Phi}+2\lambda_1 v_\Phi^2
\end{pmatrix}
\end{align}
where we defined $\lambda_{34}=\lambda_3+\lambda_4$. The masses of the light and heavy eigenstates are given as
\begin{eqnarray}
m_{h}^2&=&\frac{1}{2}[A+B-\sqrt{(A-B)^2+4C^2}], \label{eq:mh0}\\
m_{H}^2&=&\frac{1}{2}[A+B+\sqrt{(A-B)^2+4C^2}], \label{eq:mH0}
\end{eqnarray}
The lighter mass eigenstate $h$ is identified as the SM Higgs boson~\cite{Aad:2012tfa,Chatrchyan:2012ufa}. 
The two mass eigenstates $h$ and $H$ are related with the $h_\chi$, $h_\Phi$ fields through the rotation matrix $R(\alpha)$ as,
\begin{align}
\begin{pmatrix}
h_\chi \\
h_\Phi
\end{pmatrix}=R(\alpha)\begin{pmatrix}
H\\
h
\end{pmatrix}=\begin{pmatrix}
\cos\alpha & -\sin\alpha \\
\sin\alpha &  \cos\alpha 
\end{pmatrix} \begin{pmatrix}
H\\
h
\end{pmatrix},\,\,\text{with} \,\, \tan 2\alpha=\frac{2C}{A-B}
\end{align}

The mass matrix for the pseudoscalars in the basis $(\eta_\chi,\,\, \eta_\Phi)$ is given by 
\begin{align}
 \mathcal{M}_{\eta}^2=
 \begin{pmatrix}
\mu_{12}^2 \frac{v_\Phi}{v_\chi}   & -\mu_{12}^2 \\
  -\mu_{12}^2 &     \mu_{12}^2\frac{v_\chi}{v_\Phi}  \\
 \end{pmatrix}
\end{align}
One sees again that the pseudoscalar mass matrix has a zero eigenvalue corresponding to the neutral Goldstone boson $G^{0}$ eaten by the $Z$, while the physical pseudoscalar Higgs mass is given by
\begin{align}
 m_{A}^{2}=\mu_{12}^{2}\frac{v^2}{v_\Phi v_\chi}.
 \label{eq:Amass}
\end{align}
The mass eigenstates are again obtained by rotating the component fields as  
\begin{align}
\begin{pmatrix}
\eta_\chi \\
\eta_\Phi
\end{pmatrix}=R(\beta)\begin{pmatrix}
G^0\\
A
\end{pmatrix}=\begin{pmatrix}
\cos\beta & -\sin\beta \\
\sin\beta &  \cos\beta 
\end{pmatrix}\begin{pmatrix}
G^0\\
A
\end{pmatrix}\,\,
\text{with} \,\, \tan\beta=\frac{v_\Phi}{v_\chi}
\end{align}
From Eq.~\ref{eq:Amass} the pseudoscalar mass is proportional to $\mu_{12}$, which comes from the explicit lepton number violating soft term $\mu_{12}^{2}\Phi^{\dagger} \chi_L$.
Without this term in the potential, this pseudoscalar would be an unwanted doublet ``majoron'', ruled out by the measurements of the invisible decay width of the $Z$ at LEP~\cite{Zyla:2020zbs}.
Such ``majoron" would also be copiously produced in stars, leading to an astrophysical disaster.
The most straightforward way to avoid this problem is to give it a mass, through Eq~\ref{eq:Amass}~\footnote{ Another possibility is to implement the spontaneous breaking of lepton number symmetry with an ``invisible'' majoron by adding a singlet scalar~\cite{Schechter:1981cv}. Since this option was considered in~\cite{Fontes:2019uld} here we stick to the simplest linear seesaw realization with explicit \lnv}

\begin{figure}[!htbp]
	\centering
        \includegraphics[height=5.5cm,width=0.45\textwidth]{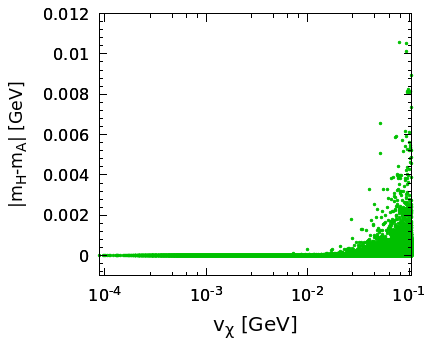}
        \includegraphics[height=5.5cm,width=0.45\textwidth]{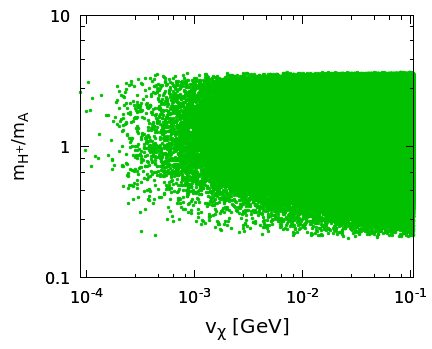}
	\caption{$|m_H-m_A|$ and $m_{H^\pm}/m_A$ versus the lepton number breaking scale $v_\chi$. Note that smallness of $v_\chi$ leads to a compressed scalar mass spectrum, see text for details. }
	\label{fig:compressed}
\end{figure}

The Higgs potential characterizing our model is rather simple.
One can describe all its quartic couplings in terms of the four physical masses, $m_h$, $m_H,m_A$ and $m_{H^\pm}$, plus the angles $\beta$ and $\alpha$. 
Indeed, the quartic couplings $\lambda_1,\lambda_2,\lambda_3$ and $\lambda_4$ can be expressed as
\begin{align}
 \lambda_1&=\frac{1}{2v^2\sin^2\beta}\Big(m_H^2\sin^2\alpha+m_h^2\cos^2\alpha-m_A^2\cos^2\beta\Big)\label{eq:lam1}\\
 \lambda_2&=\frac{1}{2v^2\cos^2\beta}\Big(m_h^2\sin^2\alpha+m_H^2\cos^2\alpha-m_A^2\sin^2\beta\Big)\label{eq:lam2}\\
 \lambda_3&=\frac{1}{v^2}\Big(2m_{H^\pm}^2-m_A^2+\frac{(m_H^2-m_h^2)\sin (2\alpha)}{\sin (2\beta)}\Big)\label{eq:lam3}\\
 \lambda_4&=\frac{2}{v^2}\Big(m_A^2-m_{H^\pm}^2\Big)\label{eq:lam4}
\end{align}
where the VEV $v = \sqrt{v^2_\Phi + v^2_\chi} = 246$~GeV $\approx v_\Phi$ and the VEV $v_\chi$ is negligibly small, see Figs.~\ref{fig:vchi} and~\ref{fig:compressed}. 

To completely specify our model let us now discuss the Yukawa couplings of the charged fermions. The most general Yukawa Lagrangian reads as 
\begin{align}
-\mathcal{L}_{\rm Yuk}=Y_e\bar{L}_L\Phi e_R + Y_u \bar{Q}_L \tilde{\Phi} u_R + Y_d \bar{Q}_L\Phi d_R + \text{H.c.}
\end{align}
Notice that charged fermions acquire mass only through their Yukawa coupling with the SM Higgs doublet $\Phi$. 
Hence, the second doublet $\chi_L$ is neutrinophilic. In this sense, this model is very close to Type-I two-Higgs-doublet Model~\cite{Grimus:2007if,Grimus:2008nb,Branco:2011iw}. \\[-.4cm]

In the mass basis the Higgs Yukawa Lagrangian can be written as
\begin{align}
-\mathcal{L}_{\rm Yuk}=\frac{m_f}{v\sin\beta}\bar{\psi}_f \psi_f \Big(\cos\alpha\, h + \sin\alpha\, H \Big),
\end{align}
for all charged fermions $f$.  

Since smallness of neutrino masses implies that the lepton number breaking VEV $v_\chi$ should be small, one finds that the Higgs boson spectrum tends to be relatively compressed.
The reason can be traced to the perturbativity constraints on the quartic couplings, in particular those on $\lambda_2$. 
As $\lambda_2$ is inversely proportional to $v_\chi^2$~(see Eq.~\ref{eq:lam2}), the numerator should be very small for small $v_\chi$ values, in order to prevent $\lambda_2$
from violating the perturbativity restrictions. \\[-.4cm]

This can only be achieved by a compressed spectrum, with the mixing angle $\beta \approx \pi/2$, while $\alpha\approx 0$ is small as required also by collider data,
  analysed in Ref.~\cite{Fontes:2019uld}. The relevant information comes from the signal strength measurements from LHC~\cite{ATLAS:2016neq,ATLAS:2019nkf}.
Indeed, in the limit $v_\chi\ll v_\Phi$, the numerator in Eq.~\ref{eq:lam2} becomes vanishingly small only if $\alpha\approx 0$ and $m_H\approx m_A$. 
This is clearly visible in Fig.~\ref{fig:compressed}.
Also note that in the limit $\alpha\approx 0$ and $m_H\approx m_A$, Eq.~\ref{eq:lam3} and Eq.~\ref{eq:lam4} can be simplified as
$\lambda_3\approx (2m_{H^\pm}^2-m_A^2)/v^2$ and $\lambda_4\approx 2(m_A^2-m_{H^\pm}^2)/v^2$. 
Hence requiring $\lambda_3$ and $\lambda_4$ to be perturbative will restrict the mass splitting between $m_{H^\pm}$ and $m_{A}$, as seen on the right panel of Fig.~\ref{fig:compressed}.

\section{$W$ Boson mass and oblique parameters $S$, $T$ and $U$} 
\label{sec:stu}

Having discussed the theoretical features of the linear seesaw model, let us now discuss the corrections to SM precision observables arising from the new particles of the model.
The oblique parameters $S$, $T$ and $U$ are very useful to parameterise the effects of new physics on electroweak observables.
Therefore we explore whether our model can provide large enough radiative corrections to the $W$-boson mass indicated by the recent CDF-II results,
while keeping the ``oblique $S$, $T$, $U$ parameters'' within their allowed 3-$\sigma$ range. 
The $S$, $T$ and $U$ parameters basically quantify the corrections to gauge boson two-point functions through loop diagrams. 
In our case, the presence of the extra scalar doublet $\chi_L$ leads to new Higgs-mediated contributions to gauge boson self-energies through loop diagrams, see Fig.~\ref{fig:SelfEnergy}
of Appendix~\ref{app:SelfEnergy}. Following the Refs.~\cite{Peskin:1991sw,Maksymyk:1993zm}, one can define the $S$, $T$ and $U$ parameters as follows:
\begin{eqnarray}
\frac{\alpha}{4 s_W^2 c_W^2}\, S &=&
\frac{A_{ZZ} \left( m_Z^2 \right) - A_{ZZ} \left( 0 \right)}{m_Z^2}
-
\left. \frac{\partial A_{\gamma \gamma} \left( q^2 \right)}
{\partial q^2} \right|_{q^2=0}
+ \frac{c_W^2 - s_W^2}{c_W s_W}
\left. \frac{\partial A_{\gamma Z} \left( q^2 \right)}
{\partial q^2} \right|_{q^2=0},
\label{S} \\[1mm]
\alpha T &=&
\frac{A_{WW} \left( 0 \right)}{m_W^2}
-
\frac{A_{ZZ} \left( 0 \right)}{m_Z^2},
\label{T} \\[1mm]
\frac{\alpha}{4 s_W^2}\, U &=&
\frac{A_{WW} \left( m_W^2 \right) - A_{WW} \left( 0 \right)}{m_W^2}
- c_W^2\,
\frac{A_{ZZ} \left( m_Z^2 \right) - A_{ZZ} \left( 0 \right)}{m_Z^2}
\no[1mm] & &
- s_W^2
\left. \frac{\partial A_{\gamma \gamma} \left( q^2 \right)}
{\partial q^2} \right|_{q^2=0}
+ 2 c_W s_W
\left. \frac{\partial A_{\gamma Z} \left( q^2 \right)}
{\partial q^2} \right|_{q^2=0},
\label{U}
\end{eqnarray}
where, $\alpha = e^2 / \left( 4 \pi \right) = g^2 s_W^2 / \left( 4 \pi \right)$ is the fine-structure constant, $s_W = \sin{\theta_W}$ and $c_W = \cos{\theta_W}$,
where $\theta_W$ is the electroweak mixing angle, and ${A}_{V V^\prime} \left( q^2 \right)$ and ${B}_{V V^\prime} \left( q^2 \right)$ are the coefficients in the vacuum-polarization tensors  
\begin{equation}
\Pi^{\mu \nu}_{V V^\prime} \left( q \right) =
g^{\mu \nu} A_{V V^\prime} \left( q^2 \right)
+ q^\mu q^\nu B_{V V^\prime} \left( q^2 \right),
\end{equation}
where $V V^\prime$ may be either $\gamma \gamma$, $\gamma Z$, $Z Z$, or $W W$, and $q = \left( q^\alpha \right)$ is the four-momentum of the gauge boson.   
The deviation of $m_W^{\rm CDF}$ from its SM prediction can be parameterized in terms of the parameters, $S$, $T$ and $U$.
The most relevant parameter is $\Delta r$, defined by the following relation~\cite{Sirlin:1980nh,Hollik:1988ii} 
\begin{align}
G_\mu = \frac{\pi \alpha}{\sqrt{2} m_W^2 s_W^2 \left( 1 - \Delta r \right)},
\label{dr}
\end{align}
where $s_W^2 \equiv 1 - m_W^2 / m_Z^2$. 
The parameter $\Delta r$ captures the structure of the loop corrections to the tree-level relation between $m_W$, $m_Z$, and the muon decay constant $G_\mu$.
It is very useful in order to determine the $W$-mass in the presence of new physics (NP). 
One can express these contributions to $\Delta r$ in terms of the $S$, $T$, and $U$ if the new fields have suppressed couplings to the light
fermions involved in the measurements of $\alpha$, $G_\mu$, $m_Z$, and $m_W$: 
\begin{align}
& \Delta r^\prime \equiv \Delta r |_\mathrm{NP} - \Delta r |_\mathrm{SM} =
\left. \frac{\partial A_{\gamma \gamma} \left( q^2 \right)}
{\partial q^2} \right|_{q^2 = 0}
+ \frac{A_{WW} \left( 0 \right) - A_{WW} \left( m_W^2 \right)}{m_W^2} \nonumber \\
& - \frac{c_W^2}{s_W^2} \left[
\frac{A_{ZZ} \left( m_Z^2 \right)}{m_Z^2}
- \frac{A_{WW} \left( m_W^2 \right)}{m_W^2} \right],
\end{align}
One then easily finds that  
\begin{align}
\Delta r^\prime = \frac{\alpha}{s_W^2} \left(
- \frac{1}{2}\, S + c_W^2 T + 
\frac{c_W^2 - s_W^2}{4 s_W^2}\, U \right).
\end{align}
This relationship can now be used to compare our model to the experimental data. 
For example, taking the measured values of $\alpha$, $G_\mu$, and $m_Z$ as experimental inputs, we can estimate the $W$ mass as
\begin{align}
m_W^2 = \left. m_W^2 \right|_\mathrm{SM}
\left( 1 + \frac{s_W^2}{c_W^2 - s_W^2}\, \Delta r^\prime \right).
\end{align}

\subsection*{$S$, $T$ and $U$ calculation}

In order to discuss the effect of the oblique $S$, $T$, $U$ parameters, we use the results of Ref.~\cite{Grimus:2007if,Grimus:2008nb,Branco:2011iw}.
To use their expressions, we must first find the matrices $U$ and $V$, which we now define explicitly~\cite{Fontes:2019uld}.
The matrix $U$ is defined as   
\begin{align}
\begin{pmatrix}
\chi^+ \\
\Phi^+
\end{pmatrix}=U\begin{pmatrix}
G^+\\
H^+
\end{pmatrix}=\begin{pmatrix}
\cos\beta & -\sin\beta \\
\sin\beta &  \cos\beta 
\end{pmatrix}\begin{pmatrix}
G^+\\
H^+
\end{pmatrix}.
\label{eq:Umatrix}
\end{align}
Hence with our convention, $U$ is same as $R(\beta)$. On the other hand $V$ is $2\times 4$ matrix defined as follows: 
\begin{align}
\begin{pmatrix}
h_\chi + i \eta_\chi \\
h_\Phi + i \eta_\Phi
\end{pmatrix} = V
\begin{pmatrix}
G^0\\
A\\
H\\
h
\end{pmatrix},\,\,\text{with}\,\, V=
\begin{pmatrix}
i R_\beta^{11}  &  i R_\beta^{12}  &  R_\alpha^{11}  & R_\alpha^{12} \\
i R_\beta^{21}  &  i R_\beta^{22}  &  R_\alpha^{21}  & R_\alpha^{22} \\
\end{pmatrix}
\label{eq:Vmatrix}
\end{align}
With the above definition of the $U$ and $V$ matrices and following the Refs.~\cite{Grimus:2007if,Branco:2011iw,Grimus:2008nb}
one can calculate the $S$, $T$, $U$ parameters~\footnote{ The symbol $U$ is used in the $S,T,U$ oblique parameters as well as to denote a matrix.}.
We have written down their explicit form in Appendix.~\ref{app:stu}.
Notice that in the limit $v_\chi\to 0$ one can take $\alpha\approx 0 $, $\beta\approx \pi/2 $, so that $S$, $T$ and $U$ take the following approximate form 
\begin{align}
T \approx \frac{1}{8 \pi s_W^2 m_W^2}  
F \left( m_{H^\pm}^2,m_H^2 \right),\,\,
S \approx \frac{1}{12\pi}\text{log}\left(\frac{m_H^2}{m_{H^\pm}^2}\right),\,\,
U \approx \frac{1}{12\pi} G\left(\frac{m_{H^\pm}^2}{m_W^2},\frac{m_H^2}{m_W^2}\right)
\label{eq:TSU}
\end{align}
where the functions $F$ and $G$ are given in Appendix.~\ref{app:stu}.  

\begin{figure}[!htbp]
	\centering
	\includegraphics[width=0.5\linewidth]{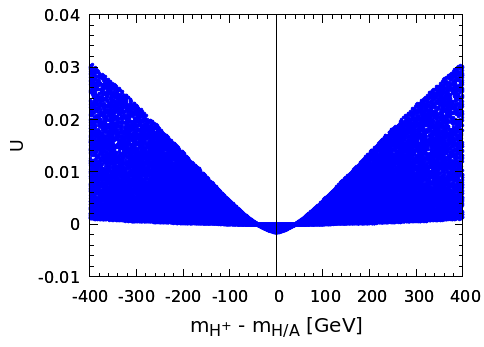}
	\caption{$U$ parameter versus $m_{H^\pm}-m_{H/A}$. Model parameters are scanned as in Table.~\ref{tab:param}.}
	\label{fig:U}
\end{figure}

On the basis of an analysis of precision electroweak data, including the latest CDF-II $W$-mass result, Ref.~\cite{Lu:2022bgw} provided the values of $S$, $T$, $U$ parameters:
\begin{align}
\big(S_0=0.06, \, \sigma_S=0.10\big), \hspace{0.5cm}
\big(T_0=0.11, \, \sigma_T=0.12\big), \hspace{0.5cm}
\big(U_0=0.14, \, \sigma_U=0.09\big)
\end{align}
with the correlation $\rho_{ST}=0.90,\,\,  \rho_{SU}=-0.59 \,\,\,\,\text{and}\,\,\,\, \rho_{TU}=-0.85$.
Here, $S_0$, $T_0$ and $U_0$ are best fit values of $S$, $T$ and $U$ respectively and $\sigma_S$, $\sigma_T$, $\sigma_U$ are the respective standard deviations.
In our model, the value of the $U$ parameter is found to be very small, see Fig.~\ref{fig:U}.
\begin{table}[!htbp]
	\begin{center}
		\begin{tabular}{|c|c| }
			\hline
			Parameter & Range \\
			\hline
			&\\[-12.5pt]
			$m_H$  & $ \,\,[150,  2000] \text{ GeV}\,\, $ \\[2.5pt]
			$m_A$  & $ [100,  2000] \text{ GeV} $ \\[2.5pt]
			$m_{H^\pm}$  & $ [100,  2000] \text{ GeV} $ \\[2.5pt]
			$v_\chi$  & $ [10^{-9},  10] \text{ GeV}$ \\[2.5pt]
			$\alpha-\beta$  & $ [-\frac{\pi}{2}, \frac{\pi}{2}] $ \\[2.5pt]
			\hline
		\end{tabular}
	\end{center}
	\caption{Value range for the numerical parameter scan for $S$, $T$ and $U$ parameters.}
	\label{tab:param}
\end{table}

\begin{figure}[!htbp]
	\centering
	\includegraphics[width=0.7\linewidth]{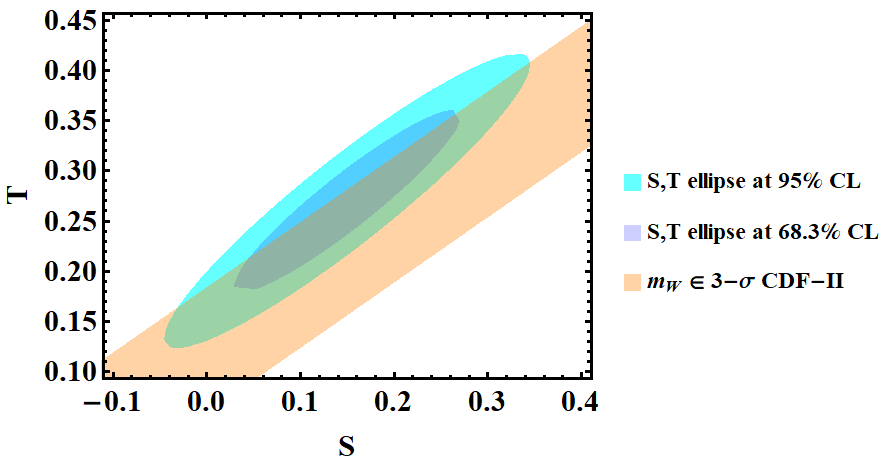}
	\caption{$S$, $T$ ellipse at various confidence levels from Eq.~\eqref{eq:STUcons}. The orange region shows the 3-$\sigma$ CDF-II measurement~\cite{CDF:2022hxs}.}
	\label{fig:stellip}
      \end{figure}

Taking $U\approx 0$, the values of $S$ and $T$ from Ref.~\cite{Lu:2022bgw} are: 
\begin{align}\label{eq:STUnonzero}
\big(S_0=0.15,\,  \sigma_S=0.08\big), \hspace{0.5cm}
\big(T_0=0.27, \, \sigma_T=0.06\big),
\end{align}
with the correlation $ \rho_{ST}=0.93 $. Hence, the $S$ and $T$ parameters are constrained as follows 
\begin{equation}
\frac{(S-S_0)^2}{\sigma_S^2} + \frac{(T-T_0)^2}{\sigma_T^2} - 2\rho_{ST}\frac{(S-S_0)(T-T_0)}{\sigma_S\sigma_T} \le R^2 (1-\rho_{ST})
\label{eq:STUcons}
\end{equation}
where $ R^2 = 2.3, 4.61, 5.99$ at $ 68.3\%, 90\%, 95\%$ confidence level~(CL).
The blue and grey ellipses in Fig. \ref{fig:stellip} follow from Eq.~\eqref{eq:STUcons} at $ 68.3\% \text{ and } 95\% $ CL.
  On the other hand the orange region shows the overlap with the CDF-II $ W $ boson mass measurement at 3-$\sigma$, showing that indeed they can be well reproduced.
  In order to find the model parameters consistent with the above constraints on $S$, $T$ we scan the parameter space as given in Table \ref{tab:param}.
For simplicity, in our numerical analysis we take just the $95\%$ CL limit for $S$ and $T$. \\
\par In addition to the $S$, $T$ restrictions,  when generating the allowed points we have also imposed following theoretical conditions: 
%
\begin{itemize}
\item We make sure that the potential is bounded from below, by imposing the vacuum stability condition given in Eq.~\eqref{vacstab}.
\item We impose perturbativity contstraints on quartic and Yukawa couplings as $\lambda_i\leq 4\pi$, $\text{Tr}(Y_\nu^\dagger Y_\nu)\leq 4\pi$ and $\text{Tr}(Y_S^\dagger Y_S)\leq 4\pi$.
\end{itemize}

We also ensure that the valid model parameter space is consistent with the neutrino physics constraints.
These include the restrictions coming from the global fit of neutrino oscillations~\cite{deSalas:2020pgw,10.5281/zenodo.4726908}
as well as from the KATRIN experiment~\cite{Aker:2021gma}, neutrinoless double beta decay searches~\cite{KamLAND-Zen:2016pfg} and cosmological data~\cite{Aghanim:2018eyx}.

In our linear seesaw scheme the neutrino mass mediators could lie in the TeV-scale accessible at high-energy colliders.
This possibility was taken up by experiments, such as ATLAS and CMS at the LHC~\cite{ATLAS:2019kpx,CMS:2018iaf,CMS:2022nty}.
Amongst the relevant collider constraints from LEP and LHC, we also impose those coming from signal strength parameter
$R_{\gamma\gamma}$~\cite{CMS:2015lsf,ATLAS:2018gfm,ATLAS:2018ntn}.

Finally, we note that in contrast to the SM, neutrinos mix~\cite{Valle:1987gv} in low-scale seesaw, even in the massless neutrino limit, so that leptonic
flavour and CP symmetries are violated,
 implying sizeable rates for \clfv processes, such as $\ell_\alpha \to \ell_\beta \gamma$, as these are not suppressed by the small neutrino
  masses~\cite{Bernabeu:1987gr,Branco:1989bn,Rius:1989gk}, for recent calculations see~\cite{Deppisch:2004fa,Deppisch:2005zm}
  \footnote{For generic \clfv references in seesaw schemes see, for example~\cite{Ilakovac:1994kj,Arganda:2007jw,Abada:2014cca,Abada:2015oba}.}.
  Detailed predictions depend, however, on Yukawa coupling matrix details, hence contraints can easily be avoided.
  
%
\begin{figure}[!h]
	\centering
    \includegraphics[width=0.507\linewidth]{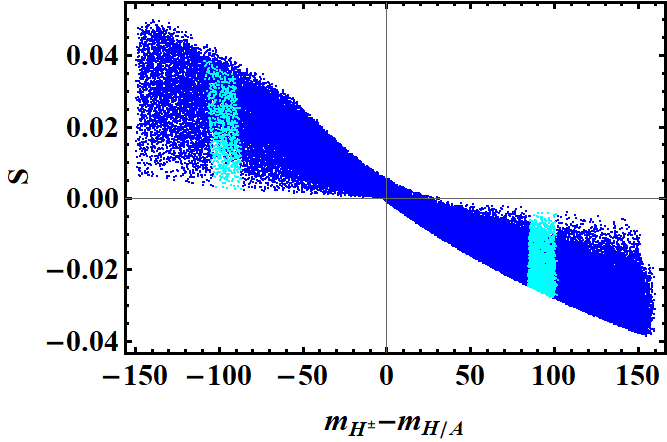}
	\includegraphics[width=0.48\linewidth]{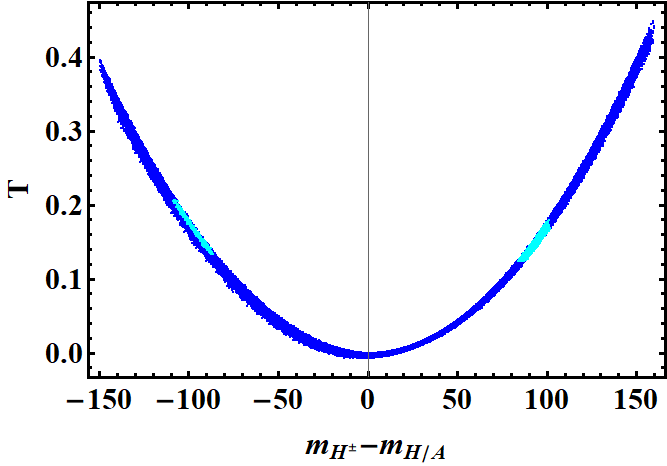}
	\includegraphics[width=0.50\linewidth]{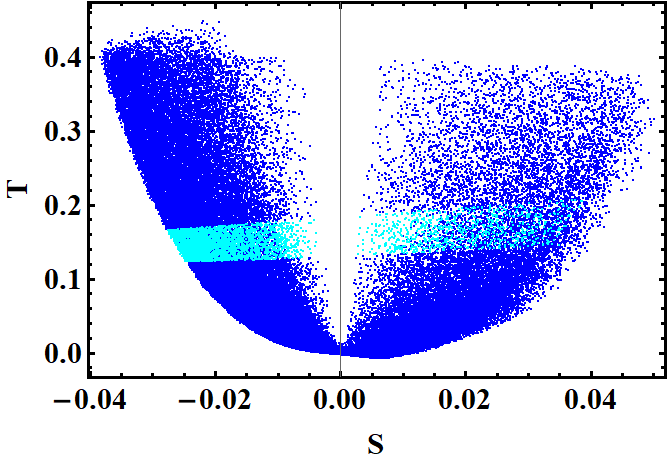}
	\caption{
          $S$, $T$ parameters versus the mass difference $m_{H^\pm}-m_{H/A}$.
          Scatter plots (from Table \ref{tab:param}) for $S$ and $T$ versus $m_{H^\pm}-m_{H/A}$ (top panels), and $T$ versus $S$ (bottom).
          The cyan points also fit the CDF result~\cite{CDF:2022hxs} at $95\%$ confidence level, according to Eq.~\eqref{eq:STUcons}. See text for details.}
	\label{fig:STU}
\end{figure}
%
\par
\begin{figure}[!ht]
	\centering
	\includegraphics[width=0.49\linewidth]{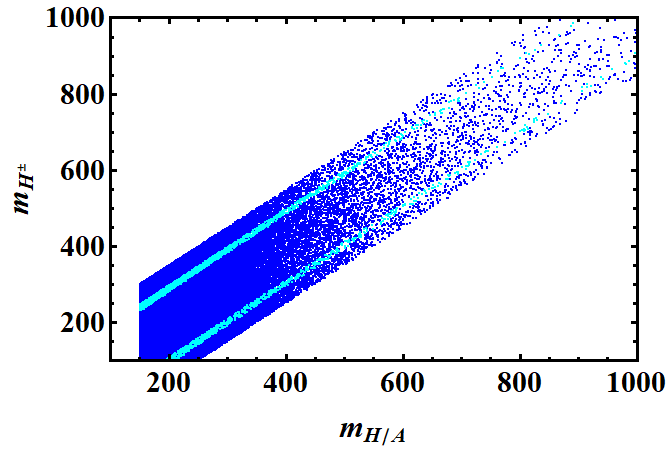}
	\caption{
          Correlation between $m_{H^\pm}$ and $m_{H/A}$ for all points given in Table \ref{tab:param}.
          The cyan points fit the CDF result~\cite{CDF:2022hxs} at $95\% $ confidence level, according to Eq.~\eqref{eq:STUcons}.}
	\label{fig:STU1}
\end{figure}

 The blue points in each panel of Fig.~\ref{fig:STU} result from imposing all of the above restrictions, except the CDF-II $W$ mass measurements.
  The cyan points are consistent both with the above restrictions as well as the CDF-II measurement.
Note that the narrow region around $S\sim 0$ with size about 0.005 is disfavored, as can be understood from Eq.~\eqref{eq:TSU}. 
On the other hand the parameter $T$ is sensitive to the mass splitting between $H^\pm$ and $H/A$, and $T$ vanishes for $m_{H^\pm}=m_{H/A}$.
In order to accommodate the $W$-mass, one need a sizable central value of the $T$ parameter, see Eq.~\eqref{eq:STUnonzero}.
As a result, one sees that $H/A$ should not be exactly degenerate in mass with $H^\pm$, as seen from Fig.~\ref{fig:STU1}. 
Note that $T$ is always positive but the sign of $S$ depends on whether $m_{H^\pm}>m_{H/A}$ or $m_{H^\pm}<m_{H/A}$.

%
\begin{figure}[!htbp]
  \centering
  \includegraphics[width=0.5\linewidth]{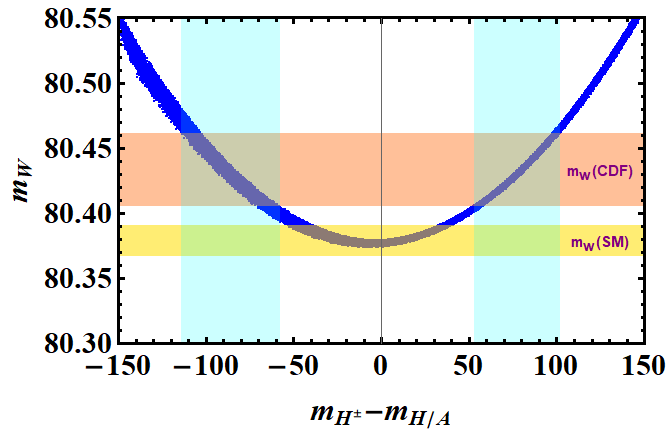}
  \caption{
    $m_W$ versus the mass difference $m_{H^\pm}-m_{H/A}$.
    The horizontal yellow band is the SM prediction, while the orange band corresponds to the 3-$\sigma$ CDF-II measurement.
    The vertical cyan bands are the regions compatible with the recent CDF-II result~\cite{CDF:2022hxs}.}
  \label{fig:last}
\end{figure}

To sum up we show in Figure \ref{fig:last} the $W$ boson mass as a function of the mass difference $m_{H^\pm}-m_{H/A}$.
  One can see that the CDF-II measurement of the $W$ mass is clearly inconsistent with the SM prediction.
  Within the SM the mass difference must be relatively small, i.e.  $m_{H^\pm}-m_{H/A} \lsim 50$~GeV.
On the other hand the $W$ boson mass is compatible with the CDF-II measurement (at 3-$ \sigma $) only when this mass difference lies in the cyan regions. 
Note that the absolute scale of the charged Higgs boson mass is not fixed, though the CDF II result suggests that it must lie below a few TeV.

\section{Conclusions}
\label{sec:conclusions}

In summary we studied the simplest linear seesaw model where a new scalar doublet $\chi_L$ with lepton number $L[\chi_L]=-2$ seeds neutrino mass generation. 
The latter is mediated by singlet quasi-Dirac leptons $N$ formed by pairs of two-component leptons $\nu_i^c$, $S_i$ with lepton numbers $L[\nu_i^c]=-1$ and $L[S_i]=1$.
The small neutrino masses follow from the tiny vacuum expectation value of the extra Higgs doublet $\chi_L$,
implying that the new neutral CP-even/CP-odd Higgs bosons $H/A$ are close in mass, as seen in the left panel of Fig.~\ref{fig:compressed}.
\par The corrections to the $W$-mass come from the loop corrections associated to the presence of the extra scalar doublet $\chi_L$
The resulting $S,T,U$ oblique parameters are shown in Figs.~\ref{fig:U},\ref{fig:stellip} and \ref{fig:STU}.
In these figures we make sure that all theoretical and phenomenological restrictions on the model are implemented.
Despite its simplicity, our well-motivated setup provides enough flexibility to explain the recent CDF-II measurement of the $W$-boson mass, 7-$\sigma$ away from the SM value.
The required charged Higgs boson masses lie in the cyan bands shown in Fig.~\ref{fig:STU1}.
One can see that the CDF-II $W$-boson mass result requires $H^{\pm}$ to have a non-zero mass splitting with $H/A$.
This is also clear from Fig.~\ref{fig:last}.

Before closing, we stress that in the linear seesaw mechanism the scale of the new fermions mediating neutrino mass generation can lie at the TeV scale,
potentially accessible to high-energy collider experiments as well as rare decay searches.
Their careful investigation, however, requires a dedicated study taking into account the detailed features of both scalar and fermion sector, outside the scope of this letter.

All in all, the simplest linear seesaw provides a simple and well-motivated setup where the recent CDF-II $W$-boson mass measurement can be in full agreement with electroweak precision tests,
neutrino physics as well as restrictions from rare \lfv decays and collider experiments.
Together with neutrino mass generation, it indicates a well-defined pattern for the new Higgs boson masses.

\begin{acknowledgments}
  The work of S.M. is supported by KIAS Individual Grants (PG086001) at Korea Institute for Advanced Study.
  The work of R.S. and A.B. is supported by the Government of India, SERB Startup Grant SRG/2020/002303.
  The work of J.V. is supported by the Spanish grants PID2020-113775GB-I00~(AEI/10.13039/501100011033) and Prometeo CIPROM/2021/054. 
\end{acknowledgments}

\appendix 
\label{Appendix}
\section{Explicit form of $S$, $T$ and $U$}
\label{app:stu}
Ref~\cite{Grimus:2007if,Grimus:2008nb,Branco:2011iw} contains the formulae for the oblique parameters in multi-Higgs-doublet models. We begin by quoting the result for $T$
\begin{align}
T &= \frac{1}{16 \pi s_W^2 m_W^2}  \Bigg\{
\sum_{b=2}^4\,
\left| \left( U^\dagger V \right)_{2b} \right|^2
F \left( m_{H^\pm}^2, \mu_b^2 \right)  - \sum_{b=2}^{3}\, \sum_{b^\prime = b+1}^4\,
\left[ \mbox{Im} \left( V^\dagger V \right)_{b b^\prime} 
\right]^2
F \left( \mu_b^2, \mu_{b^\prime}^2 \right) \nonumber \\
& + 3\, \sum_{b=2}^4\,
\left[ \mbox{Im} \left( V^\dagger V \right)_{1b} \right]^2
\left[
F \left( m_Z^2, \mu_b^2 \right) - F \left( m_W^2, \mu_b^2 \right)
\right]   - 3 \left[
F \left( m_Z^2, m_h^2 \right) - F \left( m_W^2, m_h^2 \right)
\right]  \Bigg\}
\end{align}
where the matrix $U$ and $V$ are defined in Eq.~\eqref{eq:Umatrix} and Eq.~\eqref{eq:Vmatrix}, respectively. $m_{H^\pm}$ is the mass of charged scalar and $\mu_b$ denotes the mass of the
neutral scalars, $\mu_b=\{\mu_2,\mu_3,\mu_4\}=\{m_A,m_H,m_h\}$. The function $F(x,y)$ is given as
\begin{equation}
F \left( x, y \right) = \left\{
\begin{array}{l}
\displaystyle{\frac{x + y}{2} - \frac{x y}{x - y}\, \ln{\frac{x}{y}}}
\ \Leftarrow x \neq y,
\\
0 \ \Leftarrow x = y.
\end{array}
\right.
\end{equation}
This is a non-negative function that vanishes if and only if its two arguments are equal, and it is symmetrical under the interchange of those two arguments. The oblique parameter $S$ is given as
\begin{align}
S &=
\frac{1}{24\pi} \Bigg\{
\left[ 2 s_W^2 - \left( U^\dagger U \right)_{22} \right]^2
G \left( z_{\pm}, z_{\pm} \right) + \sum_{b=2}^{3} \sum_{b^\prime = b+1}^4
\left[ \mbox{Im} \left( V^\dagger V \right)_{b b^\prime} 
\right]^2
G \left( z_b, z_{b^\prime} \right) \nonumber \\
&  - 2  \left( U^\dagger U \right)_{22} \ln{m_{H^\pm}^2}
+ \sum_{b=2}^4 \left( V^\dagger V \right)_{bb} \ln{\mu_b^2}
- \ln{m_h^2}  + \sum_{b=2}^4 \left[ \mbox{Im} \left( V^\dagger V \right)_{1b} 
\right]^2
\hat G \left( z_b \right)
- \hat G \left( z_h \right)  \Bigg\}
\end{align}
with $z_{\pm}=\frac{m_{H^\pm}^2}{m_Z^2}$, $z_{b}=\frac{\mu_{b}^2}{m_Z^2}$, $z_{h}=\frac{m_{h}^2}{m_Z^2}$. The oblique parameter $U$ is given as
\begin{align}
U &=
\frac{1}{24\pi} \Bigg\{  \sum_{b=2}^4
\left| \left( U^\dagger V \right)_{2b} \right|^2
G \left( w_{\pm}, w_b \right)  - 
\left[ 2 s_W^2 - \left( U^\dagger U \right)_{22} \right]^2
G \left( z_{\pm}, z_{\pm} \right)   \nonumber \\
& - \sum_{b=2}^{3} \sum_{b^\prime = b+1}^4
\left[ \mbox{Im} \left( V^\dagger V \right)_{b b^\prime} 
\right]^2
G \left( z_b, z_{b^\prime} \right) + \sum_{b=2}^4
\left[ \mbox{Im} \left( V^\dagger V \right)_{1b} \right]^2
\left[
\hat G \left( w_b \right)
- \hat G \left( z_b \right)
\right] \nonumber \\
& - \hat G \left( w_h \right)
+ \hat G \left( z_h \right)
 \Bigg\}
\end{align}
where $w_{\pm}=\frac{m_{H^\pm}^2}{m_W^2}$, $w_{b}=\frac{\mu_{b}^2}{m_W^2}$, $w_{h}=\frac{m_{h}^2}{m_W^2}$. The function $G(x,y)$ and $\hat G(x)$ are defined as follows
\begin{eqnarray}
G \left( x, y \right) &=& - \frac{16}{3}
+ 5 \left( x + y \right) - 2 \left( x - y \right)^2
\no & &
+ 3 \left[ \frac{x^2 + y^2}{x - y}
- x^2 + y^2 + \frac{\left( x - y \right)^3}{3} \right] \ln{\frac{x}{y}}
\no & &
+ \left[ 1 - 2 \left( x + y \right) + \left( x - y \right)^2 \right]
f \left( x + y - 1,
1 - 2 \left( x + y \right) + \left( x - y \right)^2 \right),
\label{rtycn} \\
\hat G \left( x \right) &=&
- \frac{79}{3} + 9 x - 2 x^2
+ \left( - 10 + 18 x - 6 x^2 + x^3 - 9\, \frac{x + 1}{x - 1} \right) \ln{x}
\no & &
+ \left( 12 - 4 x + x^2 \right) f \left( x, x^2 - 4 x \right),
\label{nsdep}
\end{eqnarray}
where
\begin{equation}
f \left( z, w \right) = \left\{
\begin{array}{l}
\displaystyle{\sqrt{w}\,
\ln{\left| \frac{z - \sqrt{w}}{z + \sqrt{w}} \right|}}
\ \Leftarrow w > 0,
\\*[3mm]
0 \ \Leftarrow w = 0,
\\
\displaystyle{2 \sqrt{-w}\, \arctan{\frac{\sqrt{-w}}{z}}}
\
\Leftarrow w < 0.
\end{array} \right.
\end{equation}
The function $G(x,y)$ is small for $x = y$ but it becomes sizable whenever $x$ and $y$ are quite far apart. We note that our model has a compact spectrum.
All the scalar masses except for the Higgs mass are close together.
In the limit where theses masses are large compared to $m_Z$, the function $F(x,y)$ are usually large compare to the function $G(x,y)$ and $\hat G(x)$.
This is one of the main reasons why we anticipate $T$ to be the main oblique correction.

\section{Self-Energy Diagrams}
\label{app:SelfEnergy}

\begin{figure}[h!]
	\centering
	\includegraphics[width=0.29\linewidth]{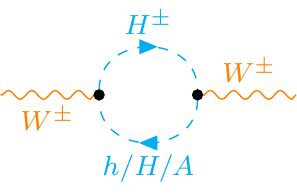}
	\includegraphics[width=0.29\linewidth]{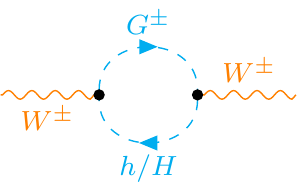}
	\includegraphics[width=0.29\linewidth]{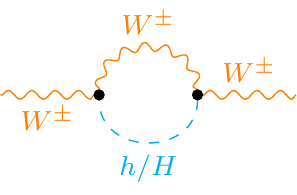}
	\includegraphics[width=0.29\linewidth]{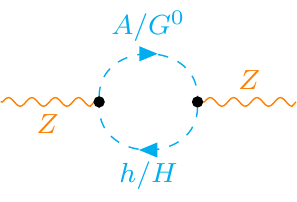}
	\includegraphics[width=0.29\linewidth]{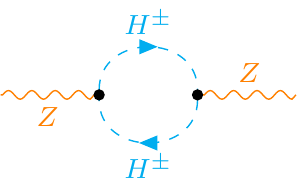}
	\includegraphics[width=0.29\linewidth]{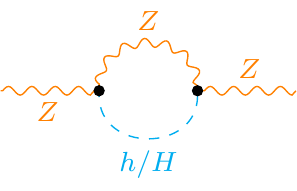}
	\includegraphics[width=0.29\linewidth]{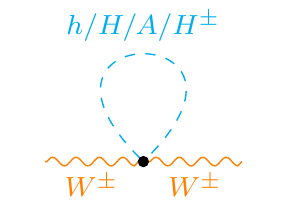}
	\includegraphics[width=0.29\linewidth]{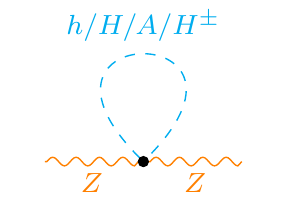}\\
	\includegraphics[width=0.29\linewidth]{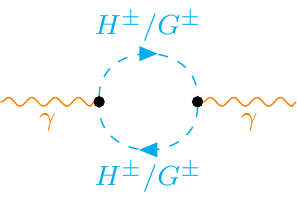}
	\includegraphics[width=0.29\linewidth]{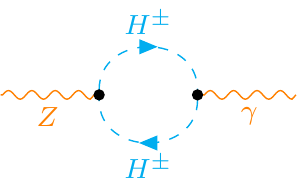}	
\caption{One-loop polarization diagrams contributing to the oblique parameters $S, T, U$.}
	\label{fig:SelfEnergy}
\end{figure}

\bibliographystyle{utphys}
\bibliography{bibliography} 

\providecommand{\href}[2]{#2}\begingroup\raggedright\begin{thebibliography}{10}

\bibitem{Aad:2012tfa}
{\bfseries ATLAS} Collaboration, G.~Aad {\em et~al.}, ``{Observation of a new
  particle in the search for the Standard Model Higgs boson with the ATLAS
  detector at the LHC},''
  \href{http://dx.doi.org/10.1016/j.physletb.2012.08.020}{{\em Phys. Lett.}
  {\bfseries B716} (2012) 1--29},
\href{http://arxiv.org/abs/1207.7214}{{\ttfamily arXiv:1207.7214 [hep-ex]}}.

\bibitem{Chatrchyan:2012ufa}
{\bfseries CMS} Collaboration, S.~Chatrchyan {\em et~al.}, ``{Observation of a
  New Boson at a Mass of 125 GeV with the CMS Experiment at the LHC},''
  \href{http://dx.doi.org/10.1016/j.physletb.2012.08.021}{{\em Phys.Lett.B}
  {\bfseries 716} (2012) 30--61},
  \href{http://arxiv.org/abs/1207.7235}{{\ttfamily arXiv:1207.7235 [hep-ex]}}.

\bibitem{Kajita:2016cak}
T.~Kajita, ``{Nobel Lecture: Discovery of atmospheric neutrino oscillations},''
  \href{http://dx.doi.org/10.1103/RevModPhys.88.030501}{{\em Rev. Mod. Phys.}
  {\bfseries 88} no.~3, (2016) 030501}.

\bibitem{McDonald:2016ixn}
A.~B. McDonald, ``{Nobel Lecture: The Sudbury Neutrino Observatory: Observation
  of flavor change for solar neutrinos},''
  \href{http://dx.doi.org/10.1103/RevModPhys.88.030502}{{\em Rev. Mod. Phys.}
  {\bfseries 88} no.~3, (2016) 030502}.

\bibitem{KamLAND:2002uet}
{\bfseries KamLAND} Collaboration, K.~Eguchi {\em et~al.}, ``{First results
  from KamLAND: Evidence for reactor anti-neutrino disappearance},''
  \href{http://dx.doi.org/10.1103/PhysRevLett.90.021802}{{\em Phys. Rev. Lett.}
  {\bfseries 90} (2003) 021802},
  \href{http://arxiv.org/abs/hep-ex/0212021}{{\ttfamily arXiv:hep-ex/0212021}}.

\bibitem{K2K:2002icj}
{\bfseries K2K} Collaboration, M.~H. Ahn {\em et~al.}, ``{Indications of
  neutrino oscillation in a 250 km long baseline experiment},''
  \href{http://dx.doi.org/10.1103/PhysRevLett.90.041801}{{\em Phys. Rev. Lett.}
  {\bfseries 90} (2003) 041801},
  \href{http://arxiv.org/abs/hep-ex/0212007}{{\ttfamily arXiv:hep-ex/0212007}}.

\bibitem{CDF:2022hxs}
{\bfseries CDF} Collaboration, T.~Aaltonen {\em et~al.}, ``{High-precision
  measurement of the $W$ boson mass with the CDF II detector},''
  \href{http://dx.doi.org/10.1126/science.abk1781}{{\em Science} {\bfseries
  376} no.~6589, (2022) 170--176}.

\bibitem{Zyla:2020zbs}
{\bfseries Particle Data Group} Collaboration, P.~Zyla {\em et~al.}, ``{Review
  of Particle Physics},'' \href{http://dx.doi.org/10.1093/ptep/ptaa104}{{\em
  PTEP} {\bfseries 2020} no.~8, (2020) 083C01}. and 2021 update.

\bibitem{Isaacson:2022rts}
J.~Isaacson, Y.~Fu, and C.~P. Yuan, ``{ResBos2 and the CDF W Mass
  Measurement},'' \href{http://arxiv.org/abs/2205.02788}{{\ttfamily
  arXiv:2205.02788 [hep-ph]}}.

\bibitem{Gao:2022wxk}
J.~Gao, D.~Liu, and K.~Xie, ``{Understanding PDF uncertainty on the $W$ boson
  mass measurements in CT18 global analysis},''
  \href{http://arxiv.org/abs/2205.03942}{{\ttfamily arXiv:2205.03942
  [hep-ph]}}.

\bibitem{Weinberg:1980bf}
S.~Weinberg, ``{Varieties of Baryon and Lepton Nonconservation},''
  \href{http://dx.doi.org/10.1103/PhysRevD.22.1694}{{\em Phys.Rev.D} {\bfseries
  22} (1980) 1694}.

\bibitem{Schechter:1980gr}
J.~Schechter and J.~W.~F. Valle, ``{Neutrino Masses in SU(2) x U(1)
  Theories},'' \href{http://dx.doi.org/10.1103/PhysRevD.22.2227}{{\em Phys.
  Rev. D} {\bfseries 22} (1980) 2227}.

\bibitem{Boucenna:2014zba}
S.~M. Boucenna, S.~Morisi, and J.~W.~F. Valle, ``{The low-scale approach to
  neutrino masses},'' \href{http://dx.doi.org/10.1155/2014/831598}{{\em Adv.
  High Energy Phys.} {\bfseries 2014} (2014) 831598},
  \href{http://arxiv.org/abs/1404.3751}{{\ttfamily arXiv:1404.3751 [hep-ph]}}.

\bibitem{Schechter:1981cv}
J.~Schechter and J.~W.~F. Valle, ``{Neutrino Decay and Spontaneous Violation of
  Lepton Number},'' \href{http://dx.doi.org/10.1103/PhysRevD.25.774}{{\em Phys.
  Rev. D} {\bfseries 25} (1982) 774}.

\bibitem{Mohapatra:1986bd}
R.~N. Mohapatra and J.~W.~F. Valle, ``{Neutrino Mass and Baryon Number
  Nonconservation in Superstring Models},''
  \href{http://dx.doi.org/10.1103/PhysRevD.34.1642}{{\em Phys. Rev. D}
  {\bfseries 34} (1986) 1642}.

\bibitem{Gonzalez-Garcia:1988okv}
M.~C. Gonzalez-Garcia and J.~W.~F. Valle, ``{Fast Decaying Neutrinos and
  Observable Flavor Violation in a New Class of Majoron Models},''
  \href{http://dx.doi.org/10.1016/0370-2693(89)91131-3}{{\em Phys. Lett. B}
  {\bfseries 216} (1989) 360--366}.

\bibitem{Akhmedov:1995ip}
E.~K. Akhmedov {\em et~al.}, ``{Left-right symmetry breaking in NJL
  approach},'' \href{http://dx.doi.org/10.1016/0370-2693(95)01504-3}{{\em Phys.
  Lett. B} {\bfseries 368} (1996) 270--280},
  \href{http://arxiv.org/abs/hep-ph/9507275}{{\ttfamily arXiv:hep-ph/9507275}}.

\bibitem{Akhmedov:1995vm}
E.~K. Akhmedov {\em et~al.}, ``{Dynamical left-right symmetry breaking},''
  \href{http://dx.doi.org/10.1103/PhysRevD.53.2752}{{\em Phys. Rev. D}
  {\bfseries 53} (1996) 2752--2780},
  \href{http://arxiv.org/abs/hep-ph/9509255}{{\ttfamily arXiv:hep-ph/9509255}}.

\bibitem{Malinsky:2005bi}
M.~Malinsky, J.~C. Romao, and J.~W.~F. Valle, ``{Novel supersymmetric SO(10)
  seesaw mechanism},''
  \href{http://dx.doi.org/10.1103/PhysRevLett.95.161801}{{\em Phys. Rev. Lett.}
  {\bfseries 95} (2005) 161801},
  \href{http://arxiv.org/abs/hep-ph/0506296}{{\ttfamily arXiv:hep-ph/0506296}}.

\bibitem{Dittmar:1989yg}
M.~Dittmar {\em et~al.}, ``{Production Mechanisms and Signatures of Isosinglet
  Neutral Heavy Leptons in $Z^0$ Decays},''
  \href{http://dx.doi.org/10.1016/0550-3213(90)90028-C}{{\em Nucl.Phys.}
  {\bfseries B332} (1990) 1--19}.

\bibitem{Gonzalez-Garcia:1990sbd}
M.~Gonzalez-Garcia, A.~Santamaria, and J.~W.~F. Valle, ``{Isosinglet Neutral
  Heavy Lepton Production in $Z$ Decays and Neutrino Mass},''
  \href{http://dx.doi.org/10.1016/0550-3213(90)90573-V}{{\em Nucl.Phys.B}
  {\bfseries 342} (1990) 108--126}.

\bibitem{AguilarSaavedra:2012fu}
J.~Aguilar-Saavedra {\em et~al.}, ``{Flavour in heavy neutrino searches at the
  LHC},'' \href{http://dx.doi.org/10.1103/PhysRevD.85.091301}{{\em Phys.Rev.D}
  {\bfseries 85} (2012) 091301},
  \href{http://arxiv.org/abs/1203.5998}{{\ttfamily arXiv:1203.5998 [hep-ph]}}.

\bibitem{Das:2012ii}
S.~Das {\em et~al.}, ``{Heavy Neutrinos and Lepton Flavour Violation in
  Left-Right Symmetric Models at the LHC},''
  \href{http://dx.doi.org/10.1103/PhysRevD.86.055006}{{\em Phys.Rev.D}
  {\bfseries 86} (2012) 055006},
  \href{http://arxiv.org/abs/1206.0256}{{\ttfamily arXiv:1206.0256 [hep-ph]}}.

\bibitem{Deppisch:2013cya}
F.~F. Deppisch, N.~Desai, and J.~W.~F. Valle, ``{Is charged lepton flavor
  violation a high energy phenomenon?},''
  \href{http://dx.doi.org/10.1103/PhysRevD.89.051302}{{\em Phys.Rev.D}
  {\bfseries 89} (2014) 051302},
  \href{http://arxiv.org/abs/1308.6789}{{\ttfamily arXiv:1308.6789 [hep-ph]}}.

\bibitem{ATLAS:2019kpx}
{\bfseries ATLAS} Collaboration, G.~Aad {\em et~al.}, ``{Search for heavy
  neutral leptons in decays of $W$ bosons produced in 13 TeV $pp$ collisions
  using prompt and displaced signatures with the ATLAS detector},''
  \href{http://dx.doi.org/10.1007/JHEP10(2019)265}{{\em JHEP} {\bfseries 10}
  (2019) 265}, \href{http://arxiv.org/abs/1905.09787}{{\ttfamily
  arXiv:1905.09787 [hep-ex]}}.

\bibitem{CMS:2018iaf}
{\bfseries CMS} Collaboration, A.~M. Sirunyan {\em et~al.}, ``{Search for heavy
  neutral leptons in events with three charged leptons in proton-proton
  collisions at $\sqrt{s} =$ 13 TeV},''
  \href{http://dx.doi.org/10.1103/PhysRevLett.120.221801}{{\em Phys.Rev.Lett.}
  {\bfseries 120} (2018) 221801},
  \href{http://arxiv.org/abs/1802.02965}{{\ttfamily arXiv:1802.02965
  [hep-ex]}}.

\bibitem{CMS:2022nty}
{\bfseries CMS} Collaboration, A.~Tumasyan {\em et~al.}, ``{Inclusive
  nonresonant multilepton probes of new phenomena at s=13{\,}{\,}TeV},''
  \href{http://dx.doi.org/10.1103/PhysRevD.105.112007}{{\em Phys.Rev.D}
  {\bfseries 105} no.~11, (2022) 112007},
  \href{http://arxiv.org/abs/2202.08676}{{\ttfamily arXiv:2202.08676
  [hep-ex]}}.

\bibitem{Drewes:2019fou}
M.~Drewes and J.~Hajer, ``{Heavy Neutrinos in displaced vertex searches at the
  LHC and HL-LHC},'' \href{http://dx.doi.org/10.1007/JHEP02(2020)070}{{\em
  JHEP} {\bfseries 02} (2020) 070},
  \href{http://arxiv.org/abs/1903.06100}{{\ttfamily arXiv:1903.06100
  [hep-ph]}}.

\bibitem{Abdullahi:2022jlv}
A.~M. Abdullahi {\em et~al.}, ``{The Present and Future Status of Heavy Neutral
  Leptons},'' in {\em {2022 Snowmass Summer Study}}.
\newblock 3, 2022.
\newblock \href{http://arxiv.org/abs/2203.08039}{{\ttfamily arXiv:2203.08039
  [hep-ph]}}.

\bibitem{Fontes:2019uld}
D.~Fontes, J.~C. Romao, and J.~W.~F. Valle, ``{Electroweak Breaking and Higgs
  Boson Profile in the Simplest Linear Seesaw Model},''
  \href{http://dx.doi.org/10.1007/JHEP10(2019)245}{{\em JHEP} {\bfseries 10}
  (2019) 245}, \href{http://arxiv.org/abs/1908.09587}{{\ttfamily
  arXiv:1908.09587 [hep-ph]}}.

\bibitem{Reig:2018ztc}
M.~Reig, D.~Restrepo, J.~W.~F. Valle, and O.~Zapata, ``{Bound-state dark matter
  with Majorana neutrinos},''
  \href{http://dx.doi.org/10.1016/j.physletb.2019.01.023}{{\em Phys. Lett. B}
  {\bfseries 790} (2019) 303--307},
  \href{http://arxiv.org/abs/1806.09977}{{\ttfamily arXiv:1806.09977
  [hep-ph]}}.

\bibitem{Barreiros:2018bju}
D.~M. Barreiros, R.~G. Felipe, and F.~R. Joaquim, ``{Combining texture zeros
  with a remnant CP symmetry in the minimal type-I seesaw},''
  \href{http://dx.doi.org/10.1007/JHEP01(2019)223}{{\em JHEP} {\bfseries 01}
  (2019) 223}, \href{http://arxiv.org/abs/1810.05454}{{\ttfamily
  arXiv:1810.05454 [hep-ph]}}.

\bibitem{Mandal:2019oth}
S.~Mandal, N.~Rojas, R.~Srivastava, and J.~W.~F. Valle, ``{Dark matter as the
  origin of neutrino mass in the inverse seesaw mechanism},''
  \href{http://dx.doi.org/10.1016/j.physletb.2021.136609}{{\em Phys. Lett. B}
  {\bfseries 821} (2021) 136609},
  \href{http://arxiv.org/abs/1907.07728}{{\ttfamily arXiv:1907.07728
  [hep-ph]}}.

\bibitem{Avila:2019hhv}
I.~M. \'Avila, V.~De~Romeri, L.~Duarte, and J.~W.~F. Valle, ``{Phenomenology of
  scotogenic scalar dark matter},''
  \href{http://dx.doi.org/10.1140/epjc/s10052-020-08480-z}{{\em Eur. Phys. J.
  C} {\bfseries 80} no.~10, (2020) 908},
  \href{http://arxiv.org/abs/1910.08422}{{\ttfamily arXiv:1910.08422
  [hep-ph]}}.

\bibitem{Grimus:2007if}
W.~Grimus, L.~Lavoura, O.~M. Ogreid, and P.~Osland, ``{A Precision constraint
  on multi-Higgs-doublet models},''
  \href{http://dx.doi.org/10.1088/0954-3899/35/7/075001}{{\em J. Phys. G}
  {\bfseries 35} (2008) 075001},
  \href{http://arxiv.org/abs/0711.4022}{{\ttfamily arXiv:0711.4022 [hep-ph]}}.

\bibitem{Grimus:2008nb}
W.~Grimus, L.~Lavoura, O.~M. Ogreid, and P.~Osland, ``{The Oblique parameters
  in multi-Higgs-doublet models},''
  \href{http://dx.doi.org/10.1016/j.nuclphysb.2008.04.019}{{\em Nucl. Phys. B}
  {\bfseries 801} (2008) 81--96},
  \href{http://arxiv.org/abs/0802.4353}{{\ttfamily arXiv:0802.4353 [hep-ph]}}.

\bibitem{Branco:2011iw}
G.~C. Branco, P.~M. Ferreira, L.~Lavoura, M.~N. Rebelo, M.~Sher, and J.~P.
  Silva, ``{Theory and phenomenology of two-Higgs-doublet models},''
  \href{http://dx.doi.org/10.1016/j.physrep.2012.02.002}{{\em Phys. Rept.}
  {\bfseries 516} (2012) 1--102},
  \href{http://arxiv.org/abs/1106.0034}{{\ttfamily arXiv:1106.0034 [hep-ph]}}.

\bibitem{Bhattacharyya:2015nca}
G.~Bhattacharyya and D.~Das, ``{Scalar sector of two-Higgs-doublet models: A
  minireview},'' \href{http://dx.doi.org/10.1007/s12043-016-1252-4}{{\em
  Pramana} {\bfseries 87} no.~3, (2016) 40},
  \href{http://arxiv.org/abs/1507.06424}{{\ttfamily arXiv:1507.06424
  [hep-ph]}}.

\bibitem{Wang:2022yhm}
L.~Wang, J.~M. Yang, and Y.~Zhang, ``{Two-Higgs-doublet models in light of
  current experiments: a brief review},''
  \href{http://arxiv.org/abs/2203.07244}{{\ttfamily arXiv:2203.07244
  [hep-ph]}}.

\bibitem{Borah:2022obi}
D.~Borah, S.~Mahapatra, D.~Nanda, and N.~Sahu, ``{Type II Dirac Seesaw with
  Observable $\Delta N_{\rm eff}$ in the light of W-mass Anomaly},''
  \href{http://arxiv.org/abs/2204.08266}{{\ttfamily arXiv:2204.08266
  [hep-ph]}}.

\bibitem{Popov:2022ldh}
O.~Popov and R.~Srivastava, ``{The Triplet Dirac Seesaw in the View of the
  Recent CDF-II W Mass Anomaly},''
  \href{http://arxiv.org/abs/2204.08568}{{\ttfamily arXiv:2204.08568
  [hep-ph]}}.

\bibitem{Batra:2022org}
A.~Batra, S.~K.~A., S.~Mandal, and R.~Srivastava, ``{W boson mass in
  Singlet-Triplet Scotogenic dark matter model},''
  \href{http://arxiv.org/abs/2204.09376}{{\ttfamily arXiv:2204.09376
  [hep-ph]}}.

\bibitem{Batra:2022pej}
A.~Batra, S.~K. A, S.~Mandal, H.~Prajapati, and R.~Srivastava, ``{CDF-II $W$
  Boson Mass Anomaly in the Canonical Scotogenic Neutrino-Dark Matter Model},''
  \href{http://arxiv.org/abs/2204.11945}{{\ttfamily arXiv:2204.11945
  [hep-ph]}}.

\bibitem{CentellesChulia:2022vpz}
S.~Centelles~Chuli\'a, R.~Srivastava, and S.~Yadav, ``{CDF-II W boson mass in
  the Dirac Scotogenic model},''
  \href{http://arxiv.org/abs/2206.11903}{{\ttfamily arXiv:2206.11903
  [hep-ph]}}.

\bibitem{Chakrabarty:2022voz}
N.~Chakrabarty, ``{The muon $g-2$ and $W$-mass anomalies explained and the
  electroweak vacuum stabilised by extending the minimal Type-II seesaw},''
  \href{http://arxiv.org/abs/2206.11771}{{\ttfamily arXiv:2206.11771
  [hep-ph]}}.

\bibitem{VanLoi:2022eir}
D.~Van~Loi and P.~Van~Dong, ``{Novel effects of the $W$-boson mass shift in the
  3-3-1 model},'' \href{http://arxiv.org/abs/2206.10100}{{\ttfamily
  arXiv:2206.10100 [hep-ph]}}.

\bibitem{Joshipura:1992hp}
A.~S. Joshipura and J.~W.~F. Valle, ``{Invisible Higgs decays and neutrino
  physics},'' \href{http://dx.doi.org/10.1016/0550-3213(93)90337-O}{{\em Nucl.
  Phys. B} {\bfseries 397} (1993) 105--122}.

\bibitem{Valle:1987gv}
J.~W.~F. Valle, ``{Resonant Oscillations of Massless Neutrinos in Matter},''
  \href{http://dx.doi.org/10.1016/0370-2693(87)90947-6}{{\em Phys.Lett.B}
  {\bfseries 199} (1987) 432--436}.

\bibitem{Bernabeu:1987gr}
J.~Bernabeu {\em et~al.}, ``{Lepton Flavor Nonconservation at High-Energies in
  a Superstring Inspired Standard Model},''
  \href{http://dx.doi.org/10.1016/0370-2693(87)91100-2}{{\em Phys.Lett.B}
  {\bfseries 187} (1987) 303--308}.

\bibitem{Branco:1989bn}
G.~Branco, M.~Rebelo, and J.~W.~F. Valle, ``{Leptonic {CP} Violation With
  Massless Neutrinos},''
  \href{http://dx.doi.org/10.1016/0370-2693(89)90587-X}{{\em Phys.Lett.B}
  {\bfseries 225} (1989) 385--392}.

\bibitem{Rius:1989gk}
N.~Rius and J.~W.~F. Valle, ``{Leptonic {CP} Violating Asymmetries in $Z^0$
  Decays},'' \href{http://dx.doi.org/10.1016/0370-2693(90)91341-8}{{\em
  Phys.Lett.B} {\bfseries 246} (1990) 249--255}.

\bibitem{Forero:2011pc}
D.~Forero, S.~Morisi, M.~Tortola, and J.~W.~F. Valle, ``{Lepton flavor
  violation and non-unitary lepton mixing in low-scale type-I seesaw},''
  \href{http://dx.doi.org/10.1007/JHEP09(2011)142}{{\em JHEP} {\bfseries 09}
  (2011) 142}, \href{http://arxiv.org/abs/1107.6009}{{\ttfamily arXiv:1107.6009
  [hep-ph]}}.

\bibitem{Cordero-Carrion:2018xre}
I.~Cordero-Carri\'on, M.~Hirsch, and A.~Vicente, ``{Master Majorana neutrino
  mass parametrization},''
  \href{http://dx.doi.org/10.1103/PhysRevD.99.075019}{{\em Phys. Rev. D}
  {\bfseries 99} no.~7, (2019) 075019},
  \href{http://arxiv.org/abs/1812.03896}{{\ttfamily arXiv:1812.03896
  [hep-ph]}}.

\bibitem{Cordero-Carrion:2019qtu}
I.~Cordero-Carri\'on, M.~Hirsch, and A.~Vicente, ``{General parametrization of
  Majorana neutrino mass models},''
  \href{http://dx.doi.org/10.1103/PhysRevD.101.075032}{{\em Phys. Rev. D}
  {\bfseries 101} no.~7, (2020) 075032},
  \href{http://arxiv.org/abs/1912.08858}{{\ttfamily arXiv:1912.08858
  [hep-ph]}}.

\bibitem{deSalas:2020pgw}
P.~F. de~Salas {\em et~al.}, ``{2020 global reassessment of the neutrino
  oscillation picture},'' \href{http://dx.doi.org/10.1007/JHEP02(2021)071}{{\em
  JHEP} {\bfseries 02} (2021) 071},
  \href{http://arxiv.org/abs/2006.11237}{{\ttfamily arXiv:2006.11237
  [hep-ph]}}.

\bibitem{10.5281/zenodo.4726908}
{P. F. De Salas} {\em et~al.}, ``{Chi2 profiles from Valencia neutrino global
  fit},'' 2021.
\newblock \url{{https://doi.org/10.5281/zenodo.4726908}}.

\bibitem{ATLAS:2016neq}
{\bfseries ATLAS, CMS} Collaboration, G.~Aad {\em et~al.}, ``{Measurements of
  the Higgs boson production and decay rates and constraints on its couplings
  from a combined ATLAS and CMS analysis of the LHC pp collision data at $
  \sqrt{s}=7 $ and 8 TeV},''
  \href{http://dx.doi.org/10.1007/JHEP08(2016)045}{{\em JHEP} {\bfseries 08}
  (2016) 045}, \href{http://arxiv.org/abs/1606.02266}{{\ttfamily
  arXiv:1606.02266 [hep-ex]}}.

\bibitem{ATLAS:2019nkf}
{\bfseries ATLAS} Collaboration, G.~Aad {\em et~al.}, ``{Combined measurements
  of Higgs boson production and decay using up to $80$ fb$^{-1}$ of
  proton-proton collision data at $\sqrt{s}=$ 13 TeV collected with the ATLAS
  experiment},'' \href{http://dx.doi.org/10.1103/PhysRevD.101.012002}{{\em
  Phys. Rev. D} {\bfseries 101} no.~1, (2020) 012002},
  \href{http://arxiv.org/abs/1909.02845}{{\ttfamily arXiv:1909.02845
  [hep-ex]}}.

\bibitem{Peskin:1991sw}
M.~E. Peskin and T.~Takeuchi, ``{Estimation of oblique electroweak
  corrections},'' \href{http://dx.doi.org/10.1103/PhysRevD.46.381}{{\em Phys.
  Rev. D} {\bfseries 46} (1992) 381--409}.

\bibitem{Maksymyk:1993zm}
I.~Maksymyk, C.~P. Burgess, and D.~London, ``{Beyond S, T and U},''
  \href{http://dx.doi.org/10.1103/PhysRevD.50.529}{{\em Phys. Rev. D}
  {\bfseries 50} (1994) 529--535},
  \href{http://arxiv.org/abs/hep-ph/9306267}{{\ttfamily arXiv:hep-ph/9306267}}.

\bibitem{Sirlin:1980nh}
A.~Sirlin, ``{Radiative Corrections in the SU(2)-L x U(1) Theory: A Simple
  Renormalization Framework},''
  \href{http://dx.doi.org/10.1103/PhysRevD.22.971}{{\em Phys. Rev. D}
  {\bfseries 22} (1980) 971--981}.

\bibitem{Hollik:1988ii}
W.~F.~L. Hollik, ``{Radiative Corrections in the Standard Model and their Role
  for Precision Tests of the Electroweak Theory},''
  \href{http://dx.doi.org/10.1002/prop.2190380302}{{\em Fortsch. Phys.}
  {\bfseries 38} (1990) 165--260}.

\bibitem{Lu:2022bgw}
C.-T. Lu, L.~Wu, Y.~Wu, and B.~Zhu, ``{Electroweak Precision Fit and New
  Physics in light of $W$ Boson Mass},''
  \href{http://arxiv.org/abs/2204.03796}{{\ttfamily arXiv:2204.03796
  [hep-ph]}}.

\bibitem{Aker:2021gma}
M.~Aker {\em et~al.}, ``{First direct neutrino-mass measurement with sub-eV
  sensitivity},'' \href{http://arxiv.org/abs/2105.08533}{{\ttfamily
  arXiv:2105.08533 [hep-ex]}}.

\bibitem{KamLAND-Zen:2016pfg}
{\bfseries KamLAND-Zen} Collaboration, A.~Gando {\em et~al.}, ``{Search for
  Majorana Neutrinos near the Inverted Mass Hierarchy Region with
  KamLAND-Zen},'' \href{http://dx.doi.org/10.1103/PhysRevLett.117.082503}{{\em
  Phys. Rev. Lett.} {\bfseries 117} no.~8, (2016) 082503},
  \href{http://arxiv.org/abs/1605.02889}{{\ttfamily arXiv:1605.02889
  [hep-ex]}}. [Addendum: Phys.Rev.Lett. 117, 109903 (2016)].

\bibitem{Aghanim:2018eyx}
{\bfseries Planck} Collaboration, N.~Aghanim {\em et~al.}, ``{Planck 2018
  results. VI. Cosmological parameters},''
  \href{http://dx.doi.org/10.1051/0004-6361/201833910}{{\em Astron.Astrophys.}
  {\bfseries 641} (2020) A6}, \href{http://arxiv.org/abs/1807.06209}{{\ttfamily
  arXiv:1807.06209 [astro-ph.CO]}}.

\bibitem{CMS:2015lsf}
{\bfseries CMS} Collaboration, V.~Khachatryan {\em et~al.}, ``{Search for a
  charged Higgs boson in pp collisions at $ \sqrt{s}=8 $ TeV},''
  \href{http://dx.doi.org/10.1007/JHEP11(2015)018}{{\em JHEP} {\bfseries 11}
  (2015) 018}, \href{http://arxiv.org/abs/1508.07774}{{\ttfamily
  arXiv:1508.07774 [hep-ex]}}.

\bibitem{ATLAS:2018gfm}
{\bfseries ATLAS} Collaboration, M.~Aaboud {\em et~al.}, ``{Search for charged
  Higgs bosons decaying via $H^{\pm} \to \tau^{\pm}\nu_{\tau}$ in the
  $\tau$+jets and $\tau$+lepton final states with 36 fb$^{-1}$ of $pp$
  collision data recorded at $\sqrt{s} = 13$ TeV with the ATLAS experiment},''
  \href{http://dx.doi.org/10.1007/JHEP09(2018)139}{{\em JHEP} {\bfseries 09}
  (2018) 139}, \href{http://arxiv.org/abs/1807.07915}{{\ttfamily
  arXiv:1807.07915 [hep-ex]}}.

\bibitem{ATLAS:2018ntn}
{\bfseries ATLAS} Collaboration, M.~Aaboud {\em et~al.}, ``{Search for charged
  Higgs bosons decaying into top and bottom quarks at $\sqrt{s}$ = 13 TeV with
  the ATLAS detector},'' \href{http://dx.doi.org/10.1007/JHEP11(2018)085}{{\em
  JHEP} {\bfseries 11} (2018) 085},
  \href{http://arxiv.org/abs/1808.03599}{{\ttfamily arXiv:1808.03599
  [hep-ex]}}.

\bibitem{Deppisch:2004fa}
F.~Deppisch and J.~W.~F. Valle, ``{Enhanced lepton flavor violation in the
  supersymmetric inverse seesaw model},''
  \href{http://dx.doi.org/10.1103/PhysRevD.72.036001}{{\em Phys.Rev.D}
  {\bfseries 72} (2005) 036001},
  \href{http://arxiv.org/abs/hep-ph/0406040}{{\ttfamily arXiv:hep-ph/0406040
  [hep-ph]}}.

\bibitem{Deppisch:2005zm}
F.~Deppisch, T.~Kosmas, and J.~W.~F. Valle, ``{Enhanced mu- - e- conversion in
  nuclei in the inverse seesaw model},''
  \href{http://dx.doi.org/10.1016/j.nuclphysb.2006.06.032}{{\em Nucl.Phys.B}
  {\bfseries 752} (2006) 80--92},
  \href{http://arxiv.org/abs/hep-ph/0512360}{{\ttfamily arXiv:hep-ph/0512360
  [hep-ph]}}.

\bibitem{Ilakovac:1994kj}
A.~Ilakovac and A.~Pilaftsis, ``{Flavor violating charged lepton decays in
  seesaw-type models},''
  \href{http://dx.doi.org/10.1016/0550-3213(94)00567-X}{{\em Nucl.Phys.B}
  {\bfseries 437} (1995) 491},
  \href{http://arxiv.org/abs/hep-ph/9403398}{{\ttfamily arXiv:hep-ph/9403398
  [hep-ph]}}.

\bibitem{Arganda:2007jw}
E.~Arganda, M.~Herrero, and A.~Teixeira, ``{mu-e conversion in nuclei within
  the CMSSM seesaw: Universality versus non-universality},''
  \href{http://dx.doi.org/10.1088/1126-6708/2007/10/104}{{\em JHEP} {\bfseries
  10} (2007) 104}, \href{http://arxiv.org/abs/0707.2955}{{\ttfamily
  arXiv:0707.2955 [hep-ph]}}.

\bibitem{Abada:2014cca}
A.~Abada, V.~De~Romeri, S.~Monteil, J.~Orloff, and A.~Teixeira, ``{Indirect
  searches for sterile neutrinos at a high-luminosity Z-factory},''
  \href{http://dx.doi.org/10.1007/JHEP04(2015)051}{{\em JHEP} {\bfseries 04}
  (2015) 051}, \href{http://arxiv.org/abs/1412.6322}{{\ttfamily arXiv:1412.6322
  [hep-ph]}}.

\bibitem{Abada:2015oba}
A.~Abada, V.~De~Romeri, and A.~Teixeira, ``{Impact of sterile neutrinos on
  nuclear-assisted cLFV processes},''
  \href{http://dx.doi.org/10.1007/JHEP02(2016)083}{{\em JHEP} {\bfseries 02}
  (2016) 083}, \href{http://arxiv.org/abs/1510.06657}{{\ttfamily
  arXiv:1510.06657 [hep-ph]}}.

\end{thebibliography}\endgroup
\end{document}